\documentclass{article}

\usepackage{float}
\usepackage{ssArxiv}
\usepackage{hyperref}
\usepackage[nameinlink]{cleveref}
\usepackage{url}
\usepackage{amssymb}
\usepackage{graphicx}
\usepackage{algorithm} 
\usepackage{algpseudocode} 
\usepackage{listings}
\usepackage{array}
\usepackage{enumitem}
\usepackage{booktabs}
\usepackage{algpseudocode}
\usepackage{xcolor}
\usepackage{caption}
\usepackage{natbib}
\usepackage{subcaption}
\usepackage{ragged2e}

\newcolumntype{L}[1]{>{\RaggedRight\hspace{0pt}}p{#1}}

\makeatletter
\def\blfootnote{\gdef\@thefnmark{}\@footnotetext}
\makeatother

\title{ObSynth: An Interactive Synthesis System for Generating Object Models from Natural Language Specifications}
\author{%
\name Alex Gu \email{gua@mit.edu}\\
\name Tamara Mitrovska$^{\star}$ \email{tamaram@mit.edu} \\
\name Daniela Velez$^{\star}$ \email{dvelez@mit.edu} \\
\name Jacob Andreas \email{jda@mit.edu} \\
\name Armando Solar-Lezama \email{asolar@csail.mit.edu} \\
\addr{Massachusetts Institute of Technology, Cambridge, MA, USA} \\
}

\begin{document}
\maketitle
\blfootnote{$^{\star}$Tamara Mitrovksa and Daniela Velez contributed equally and are ordered alphabetically by last name.}
\begin{abstract}
We introduce ObSynth, an interactive system leveraging the domain knowledge embedded in large language models (LLMs) to help users design object models from high level natural language prompts. This is an example of \emph{specification reification}, the process of taking a high-level, potentially vague specification and reifying it into a more concrete form. We evaluate ObSynth via a user study, leading to three key findings: first, object models designed using ObSynth are more detailed, showing that it often synthesizes fields users might have otherwise omitted. Second, a majority of objects, methods, and fields generated by ObSynth are kept by the user in the final object model, highlighting the quality of generated components. Third, ObSynth altered the workflow of participants: they focus on checking that synthesized components were correct rather than generating them from scratch, though ObSynth did not reduce the time participants took to generate object models. 
%%%%%%%%%%%%%%%%%%%%%%%%%%%%%%%%%%%

% In addition, we design a scoring metric to evaluate the success of synthesized object models on seven sample tasks such as classroom management and pet store applications. We demonstrate that our language-model-based synthesizers generate object models that are comparable in quality to human-generated ones.
\end{abstract}

\section{Introduction}
Recent years have seen several applications of large language models (LLMs) to support software development. For example, GitHub's Copilot has demonstrated the potential of LLMs to help programmers during the development process, and AlphaCode \citep{li2022competition} has demonstrated the possibility of solving programming competition problems using LLMs. However, both of these systems focus on very local problems---writing the next few lines of code, or a single self-contained algorithm. Creating software, however, requires much more than implementing functions from well-defined specifications. In particular, an important part of software development is leveraging domain knowledge to turn high-level application requirements into a detailed description of all the components and interfaces that will make up the application.

We propose this \textbf{specification reification} task as a new challenge at the intersection of human-computer interaction and program synthesis. Specification reification is the problem of taking a high-level, potentially vague specification of a problem and reifying it into a more concrete form. For example, consider a developer who is designing a classroom management application in an object-oriented language. Existing program synthesis systems can implement specific functions in this application---for example, a function to search for students who have not submitted an assignment---but before a developer gets to that point, they first need to design the application itself. This involves deciding which objects they need, and for each object deciding on their fields and methods, and for each method deciding what its specification should be. Specification reification is about deriving this design from the high-level description of the application. Because of the vague nature of this task, it is important that systems addressing these tasks involve humans in the loop.

\begin{figure}
    \centering
    \includegraphics[width=0.95\textwidth]{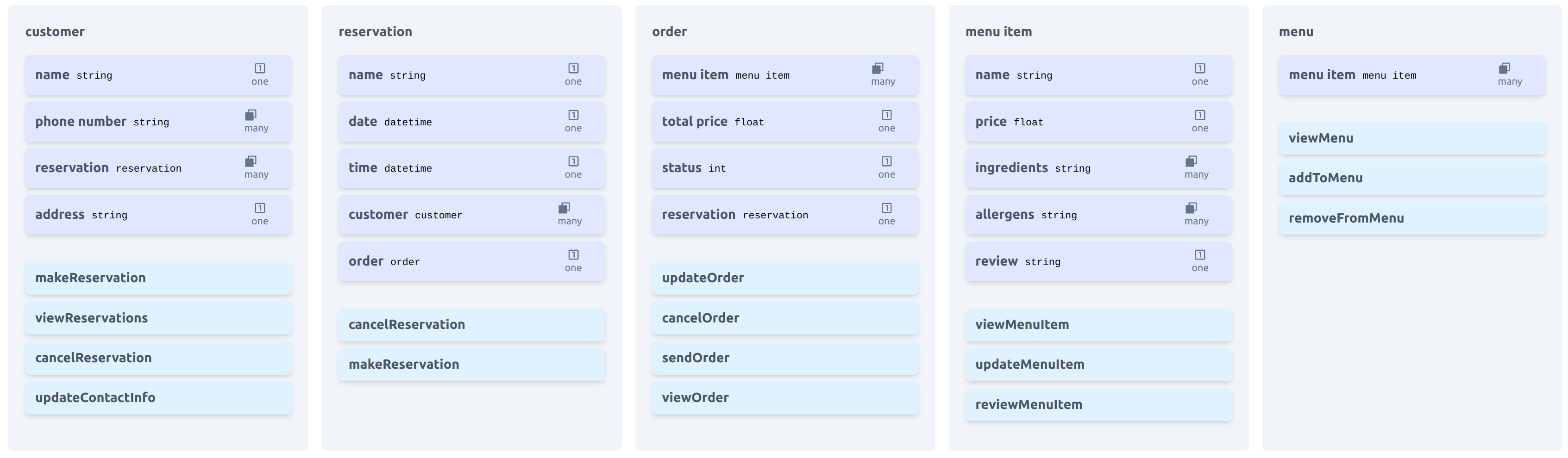}
    \caption{A sample object model for a restaurant management application, designed with the help of ObSynth. ObSynth is our interactive tool for designing object models consisting of objects, fields, types, and methods.}
    \label{fig:sample_om}
\end{figure}

In addition to introducing the new task, this paper presents \textbf{ObSynth}, a prototype interactive system for specification reification. ObSynth focuses on a key sub-task of the full specification reification problem: designing an \textbf{object model} that will make up an application from a natural language specification alone. We define an object model to be a set of objects, fields, and methods. Each field has a name and a type; types may be primitives (int, boolean, float, string, datetime), other objects, or lists of either. In our model, each method simply has a name (though we hope to extend this in future work). As an example, consider the specification \textit{``I want a restaurant management app tracking customers, their reservations, their orders, and menu items.''}. An example of an object model created via interaction with ObSynth is shown in Fig.~\ref{fig:sample_om}. In this example model, the objects are named \textit{customer, reservation, order, menu item, menu}. In Fig. \ref{fig:sample_om}, the \textit{customer} object has a field named \textit{address} of type \textit{string} and a field named \textit{reservation} of type \textit{List[reservation]}, the list type indicated by the \textit{many} icon. In Fig. \ref{fig:sample_om}, two of the \textit{customer} object's methods are named \textit{makeReservation} and \textit{updateContactInfo}.

The design of object models is challenging for a few reasons. First, the initial description is not a complete specification for the object model, so relevant pieces of the object model must be inferred. For example, in Fig. \ref{fig:sample_om}, the \textit{menu} object is not specified in the prompt, nor is the \textit{customer} object's \textit{phone number} field or \textit{makeReservation} method. The existence of these fields must be inferred by the system from its background knowledge of objects in the world and their relevant attributes. Second, the model must understand the relationship between objects, e.g., that a customer has one name and a list of reservations but no price. Even with a sophisticated language model, getting every detail right in one shot is challenging, so it is important for such a system to be designed for interactivity. This way, the tool can enhance the user's creativity rather than attempting to substitute for it. 

We present ObSynth, our synthesis-based interactive tool for solving this task. First, in Sec. \ref{sec:frontend},  we present the ObSynth UI, describing the workflow when interacting with our system. Next, in Sec. \ref{sec:backend}, we discuss in depth how we use LLM's to equip ObSynth with automation capabilities. Finally, in Sec. \ref{sec:user-study}, we discuss our user study, highlighting the ways ObSynth improves the process of creating object models. Our contributions are as follows:
\begin{enumerate}
\item We introduce and highlight a new task in program synthesis: specification reification. We also introduce the object model synthesis task as an important sub-problem of specification reification.
\item We design an interactive system, ObSynth, that assists humans in completing this task by automating parts of the process. Instead of designing the object model purely from scratch, ObSynth synthesizes a set of initial object model that the user can then build off of. Users can also ask ObSynth to automatically add relevant objects, methods, and fields at any point in the design process.
\item We conduct a user study $(n=11)$ that helps us understand how ObSynth can help participants design better object models. Through this user study, we discovered three key findings. First, object models designed using ObSynth are more detailed, showing that it often synthesizes fields users might have otherwise omitted. Second, a majority of objects, methods, and fields generated by ObSynth were kept by the user in the final object model, highlighting the quality of generated components. Third, ObSynth altered the workflow of participants: they focus on checking that synthesized components were correct rather than generating them from scratch. However, ObSynth did not reduce the time participants took to generate object models.
\end{enumerate}

\section{Related Work}
\textbf{Program Synthesis:}
The field of program synthesis has had a long history, with a variety of approaches summarized by \citet{gulwani2017program}. The first line of approaches to appear mostly focused on inductive synthesis (matching a set of input-output examples) approaches such as bottom-up search \citep{alur2015synthesis}, top-down search \citep{feser2015synthesizing}, type-directed search \citep{osera2015type}, and constraint-solving \citep{singh2016swapper}. Later, however, richer forms of program specifications were used for synthesis. 

In recent years, with new developments in machine learning, there have been more and more works exploring the potential of augmenting traditional synthesis techniques with neural networks; \citep{chaudhuri2021neurosymbolic} provides a complete survey. These include approaches to learn abstractions and libraries from scratch \citep{ellis2020dreamcoder,wong2021leveraging,nye2020learning}, execution-guided approaches that evaluate partial program states \citep{nye2020representing,gupta2020synthesize,chen2018execution}, and approaches guided by natural language information \citep{wong2021leveraging,ye2020optimal,ye2020benchmarking,nye2019learning,polosukhin2018neural}.

\textbf{Ontologies and Knowledge Graphs:}
There has also been a body of work that aims to build ontologies and knowledge graphs of natural language concepts, such as Yago ~\citep{suchanek2007yago}, WordNet ~\citep{miller1995wordnet}, and DBpedia ~\citep{auer2007dbpedia}. While these knowledge graphs have been applied in traditional NLP tasks such as question answering ~\citep{boinski2020nlp}, they are unable to provide specific insights for our synthesis task such as synthesizing fields for a certain object. As an example, when searching for nearest neighbors related to \textit{student}, WordNet comes up with synonyms such as \textit{pupil}, \textit{educatee}, and \textit{scholar}, while Yago provides a Wikipedia page for a student, a definiton of a student in Spanish, and an image containing many students. In addition, our synthesis task is very contextual: the fields of a student object would be very different if we were designing an app for teachers to manage the classroom vs a social app for students to make friends with one another. It is difficult to capture this form of context via ontologies and knowledge graphs.

%  things are very contextual
%  classroom social app vs hand in assignment app
%  wordnet 
\textbf{Large Language Models:} Recent years has also seen the birth of new works leveraging large language models (LLMs) like GPT-3 \citep{brown2020language} to perform program synthesis. A few months ago, GitHub released a powerful code autocompletion tool called \textit{GitHub Copilot} which uses context such as natural language comments and previous code in order to generate code. Copilot is built off of OpenAI's powerful machine learning model Codex \citep{chen2021evaluating}, which translates natural language to code in almost a dozen programming languages. CodeBERT ~\citep{feng2020codebert} learns representations of code and natural language for downstream tasks like code search and code documentation generation. 
\citet{heyman2021natural} use GPT-2 trained on a corpus of well-documented and commented code to synthesize programs for data science and machine learning. Building off of LLMs, ~\citet{austin2021program} incorporate human feedback to repair generated code.

There have been other works combining traditional program synthesis techniques with large language models. \citet{verbruggen2021semantic} uses traditional inductive synthesis techniques with GPT-3 to learn small intermediate functions that cannot be represented symbolically. Jigsaw ~\citep{jain2021jigsaw} uses  LLMs to synthesize code but use program analysis techniques to do post-processing. \citet{rahmani2021multi} take a component-based synthesis approach guided by LLMs which, for example, help rank candidate programs. Our approach differs from existing works in LLMs in that we approach synthesis from a global view, generating the overall structure of \emph{applications} rather than the local structure of code itself.

\section{The ObSynth Frontend}\label{sec:frontend}
\begin{figure}
    \centering
    \includegraphics[width=0.9\textwidth]{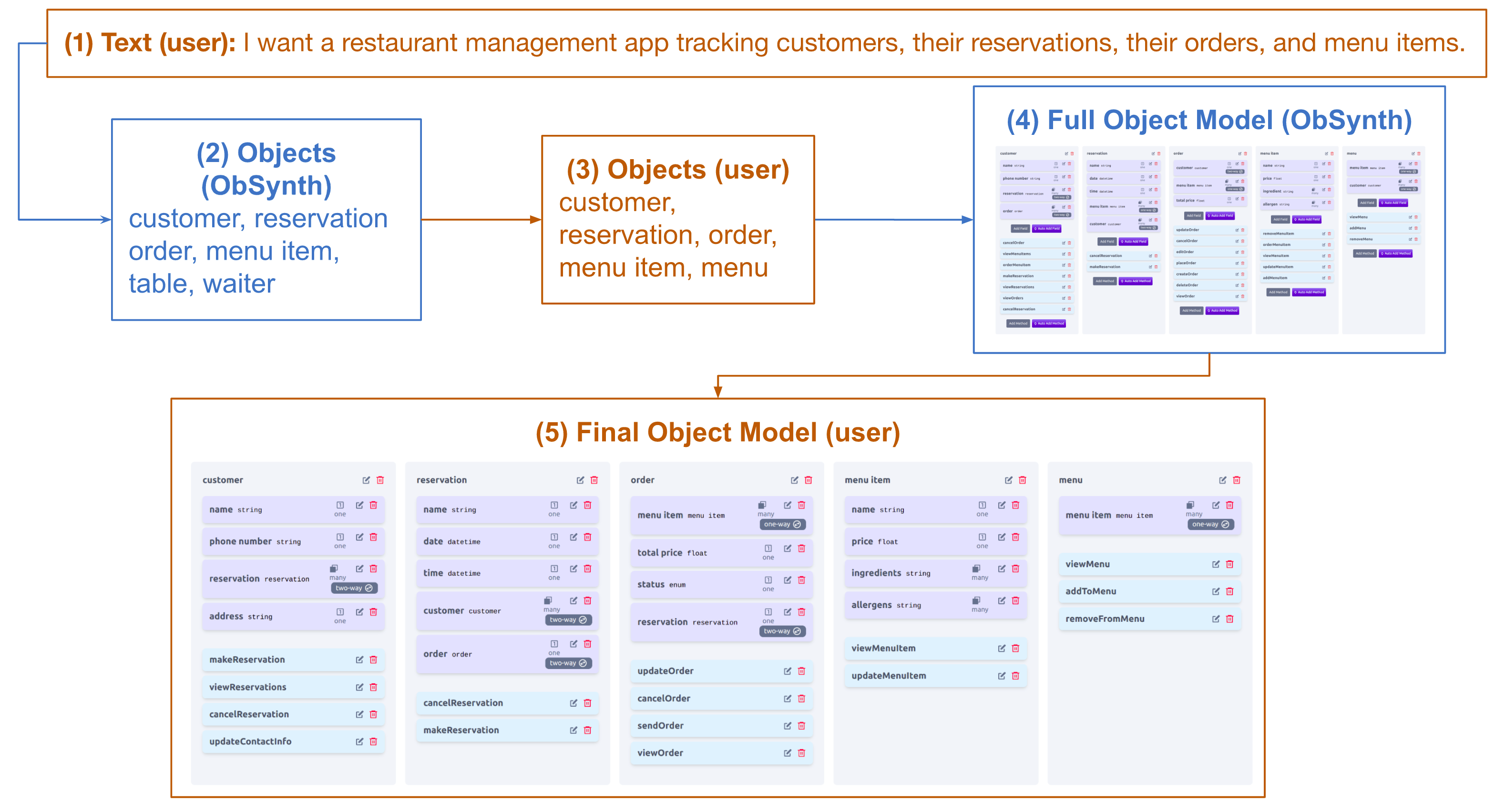}
    \caption{Complete workflow of a user synthesizing an object model from a text specification using ObSynth. Orange indicates user interaction, blue indicates ObSynth automation.}
    \label{fig:flowchart}
\end{figure}

\begin{figure}
    \centering
    \includegraphics[width=0.9\textwidth]{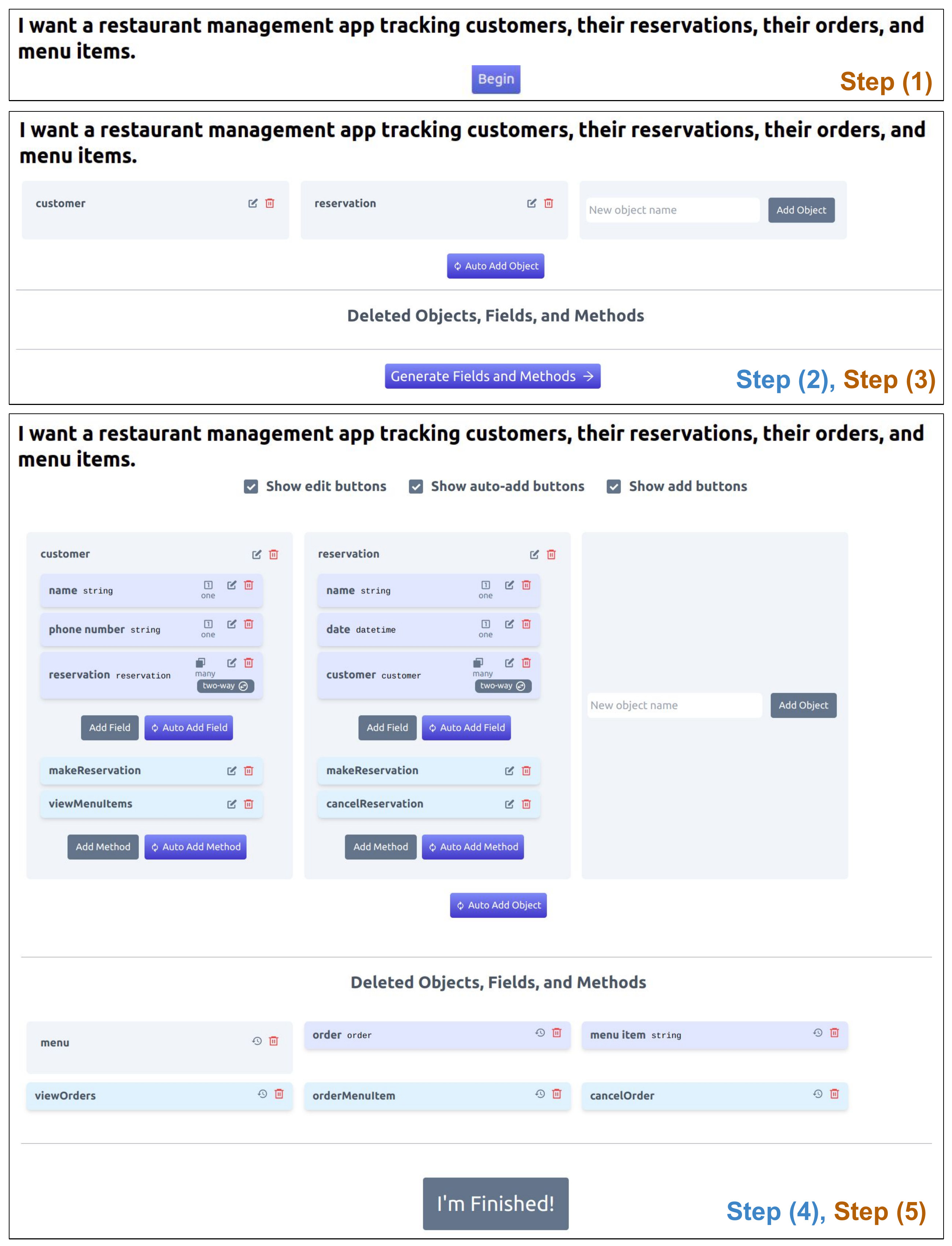}
    \caption{Full ObSynth UI. In step (1), the user specifies a prompt. In step (2), ObSynth automatically generates a few object names, and in step (3) the user can freely edit these with the potential help of the \textit{Auto Add Object} synthesis feature. Next, in step (4), ObSynth synthesizes fields, types, multiplicities, and methods for each specified object (full model not shown due to space). Finally, in step (5), the user can edit the synthesized model (with automated synthesis features) to generate a desired final object model.}
    \label{fig:ui-full}
\end{figure}

In this section, we describe our vision of how users interact with ObSynth to generate a final specification from a text prompt. Fig. \ref{fig:flowchart} shows the steps of the ObSynth workflow at a high level, while Fig. \ref{fig:ui-full} shows the concrete UI users work with at each step. In Sec. \ref{subsec:text-initial}, we explain steps (1)-(4), where the user specifies an initial text prompt and works with ObSynth to obtain an initial full object model. Then, in Sec. \ref{subsec:initial-final}, we explain step (5), where users use ObSynth to tweak this object model to fit their use case. 

\subsection{From Text to Initial Object Model}\label{subsec:text-initial}
The user starts by specifying a text prompt as shown in Fig. \ref{fig:ui-full}, step (1). As a running example, we use the prompt \textit{``I want a restaurant management app tracking customers, their reservations, their orders, and menu items.''} When the user enters this prompt, ObSynth synthesizes a list of object names without fields or methods---in this case \textit{customer, reservation, order, menu item, table, waiter}, as shown in Fig. \ref{fig:ui-full}, step (2). The user can then add, edit, or delete these object names in the same UI. ObSynth also provides an additional functionality, which will attempt to synthesize additional relevant objects (the purple \textit{Auto Add Object} button). This feature could potentially suggest objects that the user may not have thought of themselves. After specifying a list of object names, ObSynth will automatically synthesize a set of fields, types, and method names for each object, as seen in step (4). This generates an initial object model specification for the text prompt.

\subsection{From Initial Object Model to Final Specification}\label{subsec:initial-final}
Since different users may have different use cases, ObSynth gives users the flexibility to edit the object model as desired. After generating a full object model, users see the UI shown in the bottom panel of Fig. \ref{fig:ui-full}. We first explain the UI features (grey buttons), and then move to the synthesis features (blue buttons). 

\textit{ObSynth UI features: } As shown in Fig. \ref{fig:ui-full}, ObSynth's UI allows users to easily add, delete, and edit their own objects, fields, and methods at all stages of the process. Deleted objects, fields, and methods can also be restored. Users can easily toggle the multiplicity of an object field by clicking the \texttt{one/many} button. ObSynth also has a \texttt{one-way/two-way} button that adds reverse object-field relationships: if a \texttt{student} object has a \texttt{teacher} field and there is a \texttt{teacher} object, the button will toggle whether the \texttt{teacher} object has a \texttt{student} field. Finally, ObSynth ensures that when the user changes the name of an object, all other fields having that object type are renamed. All these buttons are shown in grey in Fig. \ref{fig:ui-full}.

\textit{ObSynth synthesis features: } However, what makes ObSynth unique is its synthesis capabilities, shown in the blue buttons in Fig. \ref{fig:ui-full}. When the user clicks \textit{Begin}, a set of initial object names is automatically synthesized for them. When the user clicks \textit{Generate Fields and Methods}, fields and methods for each object are likewise automatically populated. The blue \textit{Auto Add} buttons allow users to synthesize specific parts of the object model: \textit{Auto Add Field} synthesizes a new field, type, and multiplicity for the current object. \textit{Auto Add Method} synthesizes a new method name for an existing object. \textit{Auto Add Object} synthesizes a new relevant object and fully populates it with fields (including types and multiplicity) and method names. These synthesis tools ease the user in the development of a suitable object model for their use case, especially giving the user ideas for objects, fields, and methods they might have overlooked.

% \begin{figure}
%     \centering
%     \includegraphics[width=0.9\textwidth]{step1.png}
%     \caption{Caption}
%     \label{fig:step1}
% \end{figure}

% \begin{figure}
%     \centering
%     \includegraphics[width=0.9\textwidth]{step1_modified.png}
%     \caption{Caption}
%     \label{fig:step1_modified}
% \end{figure}

% \begin{figure}
%     \centering
%     \includegraphics[width=0.9\textwidth]{fields.png}
%     \caption{ObSynth UI, showing the objects, fields, types, multiplicities, and methods for a classroom application synthesized from object names}
%     \label{fig:fields}
% \end{figure}
\section{The ObSynth Backend} \label{sec:backend}
In Sec. \ref{sec:frontend}, we described the frontend of ObSynth and how users can interact with the synthesis features of ObSynth to design an object model. In this section, we go into the precise technical details of how these synthesis components work.

\subsection{Large Language Models and GPT-3 Prompt Engineering}
In recent years, large language models (LLMs) trained on enormous amounts of data have shown new capabilities on natural language tasks. One such model is GPT-3, which is a 175B parameter model trained on about 45 TB of data. Synthesizing object models require external knowledge beyond the initial text specification. Because GPT-3 has  knowledge of the real world, we leverage this knowledge via a method known as \textit{prompt engineering}. 

The premise of prompt engineering is that by specifying examples of a task we wish GPT-3 to perform, it can learn to perform the task in a few-shot manner. For example, one task is to synthesize initial object names from a specification. In this subtask, described at length in Sec. \ref{subsec:end-to-end}, we give it an example of a classroom management application with initial object names synthesized and ask it to do the same task for other applications. This is the main technique that powers ObSynth's synthesis functionality.

ObSynth uses the GPT-3 Q\&A API with the text-davinci-001 model and default parameters from the OpenAI web API\footnote{https://beta.openai.com/playground}: a temperature of $0$ (greedy decoding), frequency penalty of $0$, and presence penalty of $0$. 

\subsection{Individual Prompt-Engineering Subtasks}
ObSynth has a few automation components to aid the user at each step of the design process, namely:
\begin{enumerate}
\item [(T1)] Synthesizing object names from an initial natural language specification (Fig. \ref{fig:flowchart}, Step 2)
\item [(T2)] Synthesizing more object names given a set of existing object names (Step 3)
\item [(T3)] Synthesizing a full object model from a set of object names (Step 4)
\item [(T4)] Synthesizing more objects, fields, and method names in an existing object model (Step 5) 
\end{enumerate}

We further break down tasks (T1)-(T4) into eight individual subtasks, each of which can be solved with a single GPT-3 prompt. In Table \ref{table:subtasks}, we describe the individual subtasks, along with the inputs and outputs we use to solve them. Since GPT-3 is a text-to-text system, we must convert back and forth between our structured object model representation and descriptive text representations. For example, in (ST1), GPT-3 gives us outputs of the form \textit{``It has the following tables: student, assignment, teacher.''}, but we must parse this to understand that this represents three object names: student, assignment, and teacher. In (ST2), we feed GPT-3 our object model's name descriptively: \textit{``The app has the following tables: student, submission, assignment, teacher.''}. We defer the presentation of an end-to-end description of a subtasks in Sec. \ref{subsec:end-to-end}.
% \begin{enumerate}
% \item [(ST1)] Synthesize object names from an initial natural language prompt 
% \item [(ST2)] Synthesize more object names from a prompt and existing object names
% \item [(ST3)] Synthesize a set of field names from a prompt and given object name
% \item [(ST4)] Predict the type of a field
% \item [(ST5)] Predict the multiplicity of a field
% \item [(ST6)] Synthesize a set of method names for a given object
% \item [(ST7)] Synthesize more field names 
% \item [(ST8)] Synthesize more method names from a given object
% \end{enumerate}
\begin{table}
\caption{Complete list of subtasks used in designing the automation features of ObSynth. Each of these subtasks requires real world knowledge and can be done via a single GPT-3 prompt.}
\begin{center}
\footnotesize
\begin{tabular}{ |L{0.35\textwidth}|L{0.42\textwidth}|L{0.17\textwidth}| } 
 \hline
 \textbf{Subtask} & \textbf{Inputs} & \textbf{Outputs} \\ 
 \hline
 (ST1) synthesize initial object names & prompt & object names \\
 \hline
 (ST2) synthesize additional object names & prompt, object names & new object names \\
 \hline
 (ST3) synthesize field names & prompt, object name & field names \\
 \hline
 (ST4) predict the type of a field & prompt, field name, all object names & field type \\
 \hline
 (ST5) predict the multiplicity of a field & prompt, object name, field name & field multiplicity \\
 \hline
 (ST6) synthesize a set of method names & prompt, object name, field names & method names \\
 \hline
 (ST7) synthesize additional field names & prompt, object name, field names & new field names \\
 \hline
 (ST8) synthesize additional method names & prompt, object name, field names, method names & new method names\\
 \hline
\end{tabular}
\end{center}
\label{table:subtasks}
\end{table}

Mapping UI buttons to tasks, the initial \textit{[Begin]} button corresponds to (T1), the \textit{[Auto Add Object]} button in the initial phase corresponds to (T2), the \textit{[Generate Objects, Methods, and Fields]} button corresponds to (T3), and the \textit{[Auto Add Object/Method/Field]} buttons correspond to (T4). Each of these buttons is a combination of several subtasks, as shown in Table \ref{table:buttons}.

\begin{table}
\caption{Subtasks called for each of the automation buttons of ObSynth. These synthesis features make ObSynth powerful, giving it the potential to synthesize objects, fields, and methods that the user may not have considered including.}
\begin{center}
\footnotesize
\begin{tabular}{ |L{0.4\textwidth}|L{0.3\textwidth}| } 
 \hline
 \textbf{Button} & \textbf{Subtasks Called} \\ 
 \hline
 (T1) Begin & (ST1) \\
 \hline
 (T2) Auto Add Object (before fields/methods) & (ST2) \\
 \hline
 (T3) Generate Fields and Methods & (ST3), (ST4), (ST5), (ST6) \\
 \hline
 (T4) Auto Add Object (after fields/methods) & (ST2), (ST3), (ST4), (ST5), (ST6) \\
 \hline
 (T4) Auto Add Field & (ST7), (ST4), (ST5) \\
 \hline
 (T4) Auto Add Method & (ST8) \\
 \hline
\end{tabular}
\end{center}
\label{table:buttons}
\end{table}

\subsection{End-to-end Pipeline of a Subtask} \label{subsec:end-to-end}
\begin{figure}
    \centering
    \includegraphics[width=0.9\textwidth]{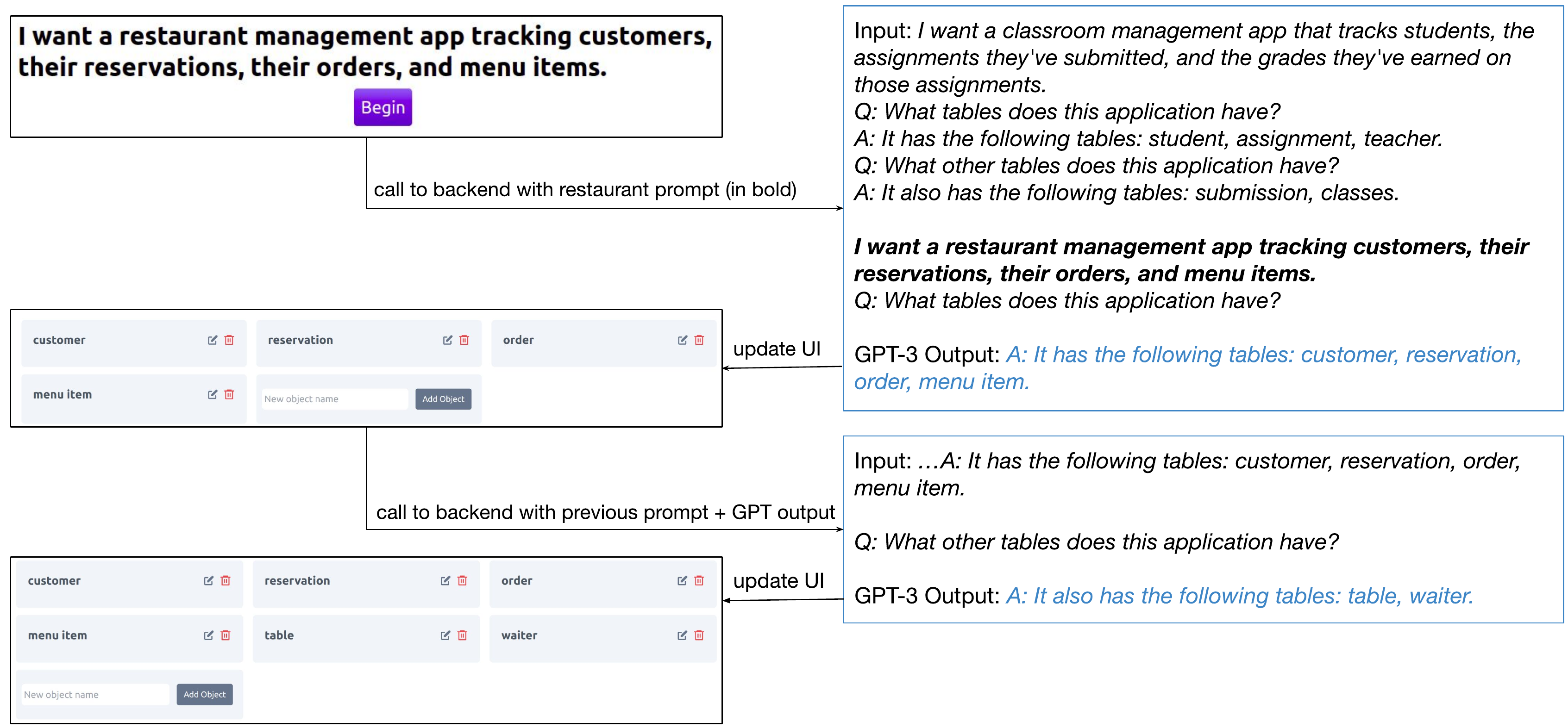}
    \caption{Interaction between frontend UI and backend GPT-3 calls for (ST1), generating a set of initial object names from a text prompt. We make use of GPT-3's real world knowledge to synthesize relevant object names.}
    \label{fig:backend-calls}
\end{figure}

In this section, we give a complete description of the prompt engineering approach used to perform (ST1). As other tasks are similar, we defer the methodology behind the rest of the subtasks to the supplementary material. (ST1) is the first automation component of ObSynth, generating initial objects from a high-level specification. Once again, we consider the restaurant management application. The interaction between the frontend and backend is shown in Fig. \ref{fig:backend-calls}. The prompt we employ works as follows: first, we give GPT-3 a full example (primer) of how it should respond for a classroom management application prompt. Then, we feed it the user-specified prompt (the restaurant management application) and ask it to generate relevant tables. Fitting the format of our primer, it responds with \textit{``A: It has the following tables: customer, reservation, order, menu item.''}, which we can parse to extract the object names. We then add those object names to the UI. Next, in order to generate a larger set of object names, we query GPT-3 a second time, using the previous prompt, the previous answer, and the new question \textit{``Q: What other tables does this application have?''}. This gives us more object names, and again extract them to the UI. In (ST1), since users are able to delete tables easily, we err on the side of overgeneration and show the user all tables produced by the two queries.

\section{User Study} \label{sec:user-study}

We conducted a user study to test if ObSynth could help users build better object models. Specifically, the goal of our user study was to answer the following four questions. 
\begin{enumerate}
    \item [(Q1)] Are object models produced with the help of ObSynth more detailed than those without?
    \item [(Q2)] Are the fields and methods in initial object models produced by ObSynth better than those without?
    \item [(Q3)] Do people using ObSynth take less time to design object models?
    \item [(Q4)] How does ObSynth shift the workflow of participants building object models?
\end{enumerate}

\subsection{User Study Design}
\begin{figure}[]
    \centering
    \includegraphics[width=0.9\textwidth]{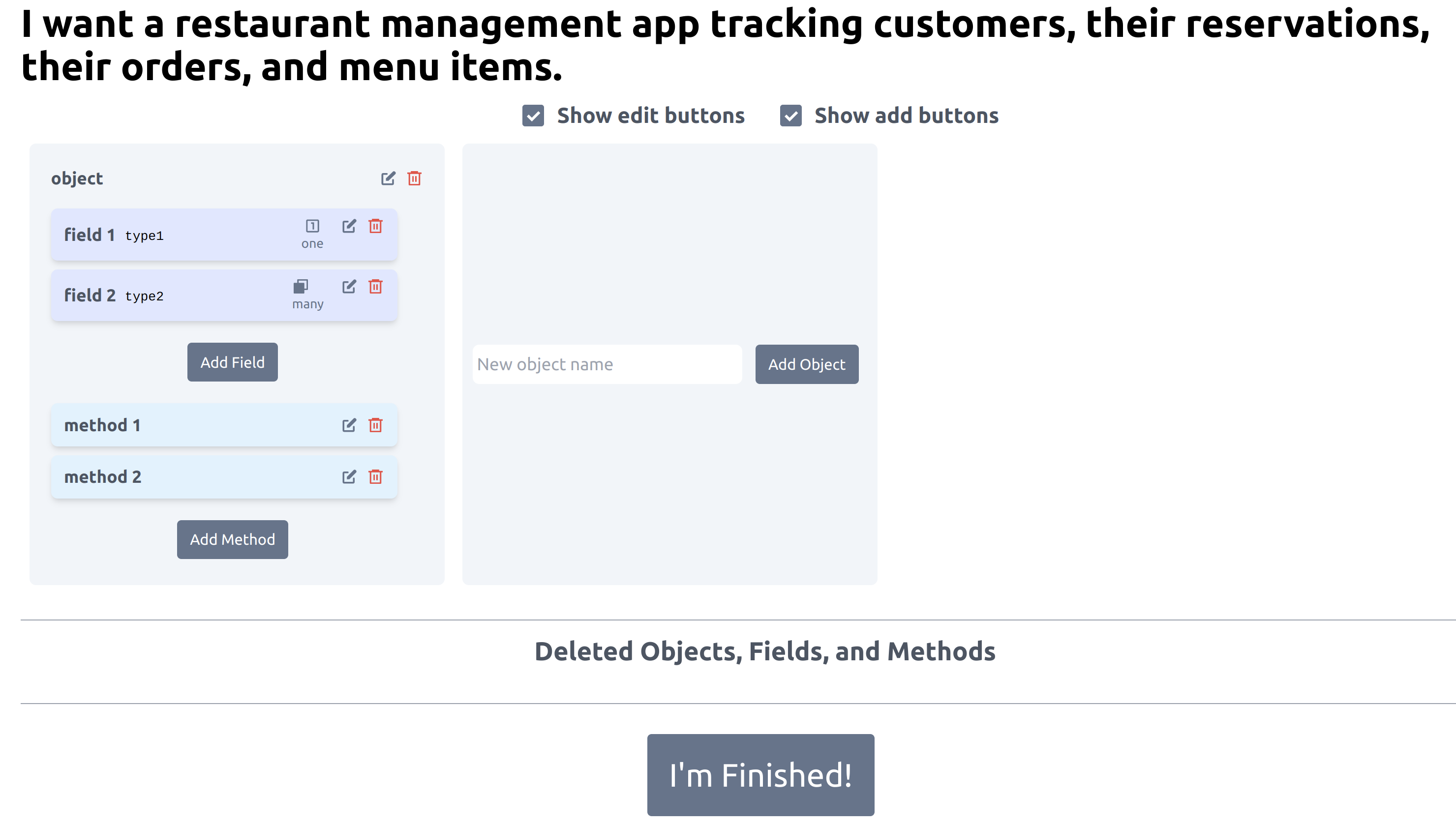}
    \caption{The UI for the control group, which is the same as the UI for ObSynth but with the synthesis features (\textit{Auto Add} buttons, \textit{Generate Fields and Methods}) removed.}
    \label{fig:ui-control}
\end{figure}

In total, 11 participants participated in our study, 6 from academia and 5 from industry. All participants were familiar with web development and object model design. We gave all participants the same prompt: \textit{``I want a restaurant management app tracking customers, their reservations, their orders, and menu items.''} Participants were split into two groups: a control group and an ObSynth group. In both groups, participants were given the same high-level instructions: to design an object model for the restaurant management application with objects, fields, types, multiplicities, and method names. In the ObSynth group, participants were shown and taught to use the ObSynth interface (Fig. \ref{fig:ui-full}). In the control group, participants were shown and taught to use the same interface with synthesis capabilities removed (Fig. \ref{fig:ui-control}). In order to better understand participant behavior, we kept track of what buttons they pressed at each time step and how the object models evolved over time. Finally, after submitting their final object models, for both groups, the participants were asked to describe their experience interacting with the system and to explain their thought process while designing the object models.

\begin{figure}[!t]
\centering
\begin{subfigure}{\textwidth}
  \centering
  \includegraphics[width=\linewidth]{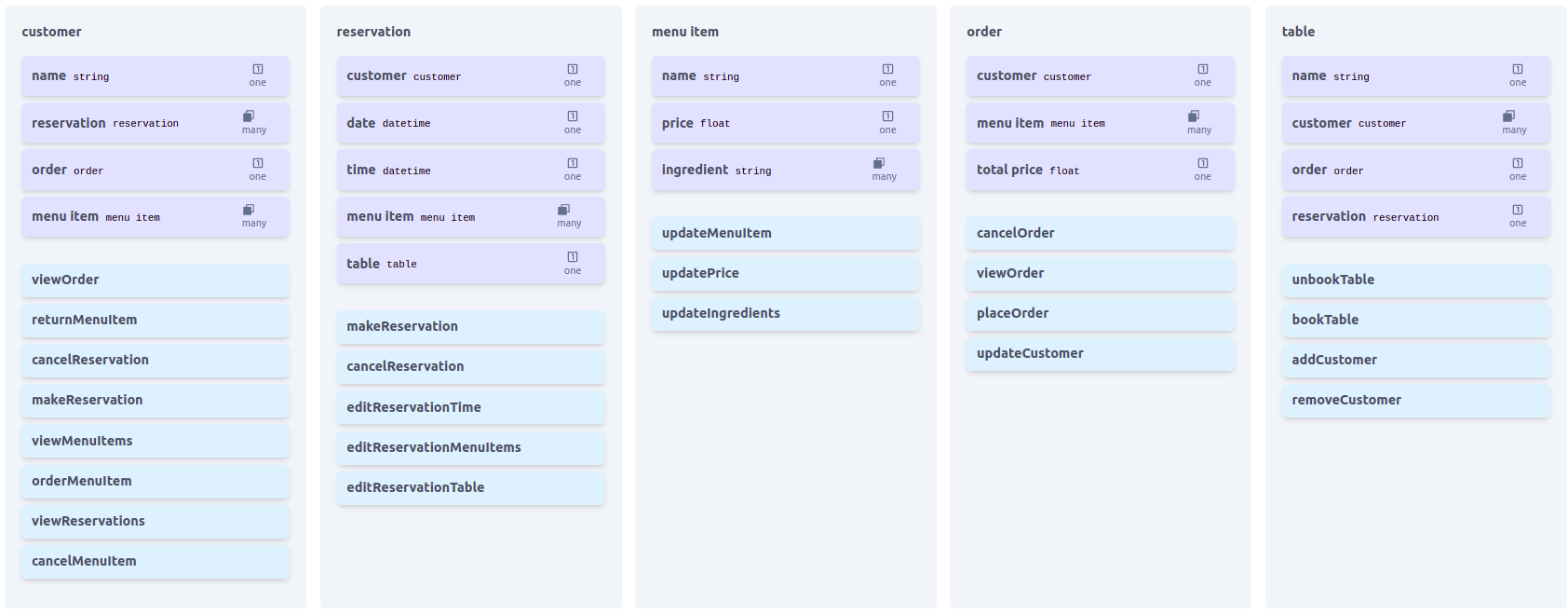}
  \caption{Object model produced with the help of ObSynth}
  \label{fig:comparison-obsynth}
\end{subfigure}
\hfill
\begin{subfigure}{\textwidth}
  \centering
  \includegraphics[width=\linewidth]{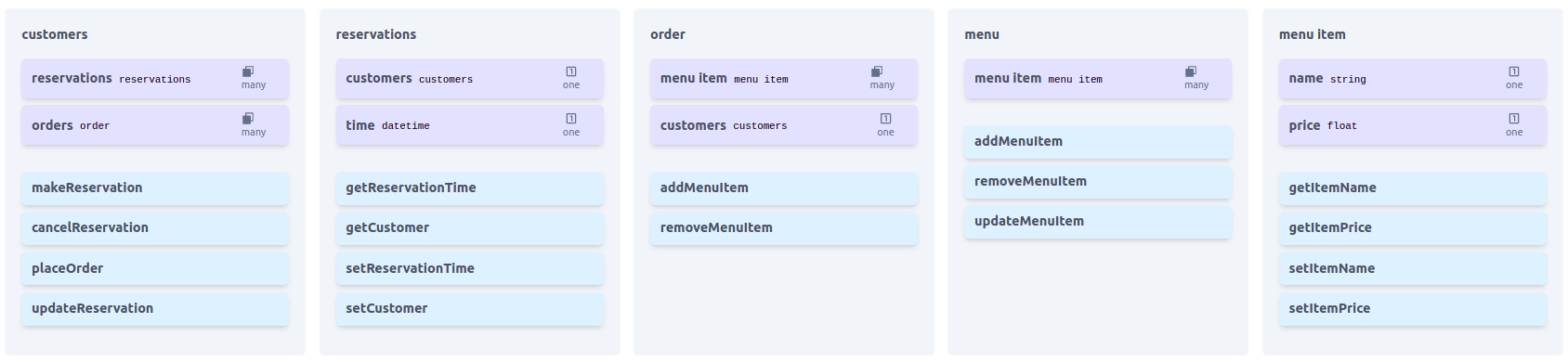}
  \caption{Object model produced without ObSynth}
  \label{fig:comparison-control}
\end{subfigure}
\caption{Comparison of two final object models produced with (top) and without (bottom) ObSynth for a restaurant management application. Observe that while both models adequately fit the prompt, the ObSynth model contains many fields and methods omitted in the control group model.}
\label{fig:comparison}
\end{figure}

In Fig. \ref{fig:comparison}, we show an example of object model produced by the ObSynth group (Fig. \ref{fig:comparison-obsynth}) and the control group (Fig. \ref{fig:comparison-control}). In the reminder of this section, we analyze the results of the user study in depth from the perspective of our four research questions.

\subsection{(Q1) Are object models produced with the help of ObSynth more detailed than those without?} 

For both the control group and the ObSynth group, users were asked to edit the models until they were satisfied that they were good models for the given prompt. However, this does not necessarily mean that all models generated by either group are equally good. On the one hand, some fields may be important for the task, but it may simply not occur to some users to include them. On the other hand, it is well known that automation can lead some users to be complacent, which in our case could mean blindly accepting the machine suggestions and losing initiative to further develop the machine generated object models, documented in a different context in \citet{LevyASS21}).
In either case, there is strong reason to believe that the main shortcoming of the resulting models would be that they miss relevant fields, since a blatantly incorrect field would call the user's attention in a way that a missing field would not. 

This assumption suggests that a way to evaluate the models produced by the system is to track which object fields are present in object models generated with the help of ObSynth vs. which are present in object models generated in the baseline setup. Fields that are present only rarely in models in both cases can be regarded as idiosyncratic, but fields that are present consistently in the control group but only rarely in the ObSynth object models would suggest a complacency effect. These would be fields that unaided users consistently regarded as important, but that users of ObSynth failed to consider. By contrast, fields that are present consistently in the ObSynth group but only rarely in the control group are suggestive of fields that did not occur to the unaided users, but the ObSynth users found them useful either after they were suggested by ObSynth, or after thinking of them themselves inspired by the suggestions given by the tool.

\begin{figure}[!t]
\centering
\begin{subfigure}{.5\textwidth}
  \centering
  \includegraphics[width=\linewidth]{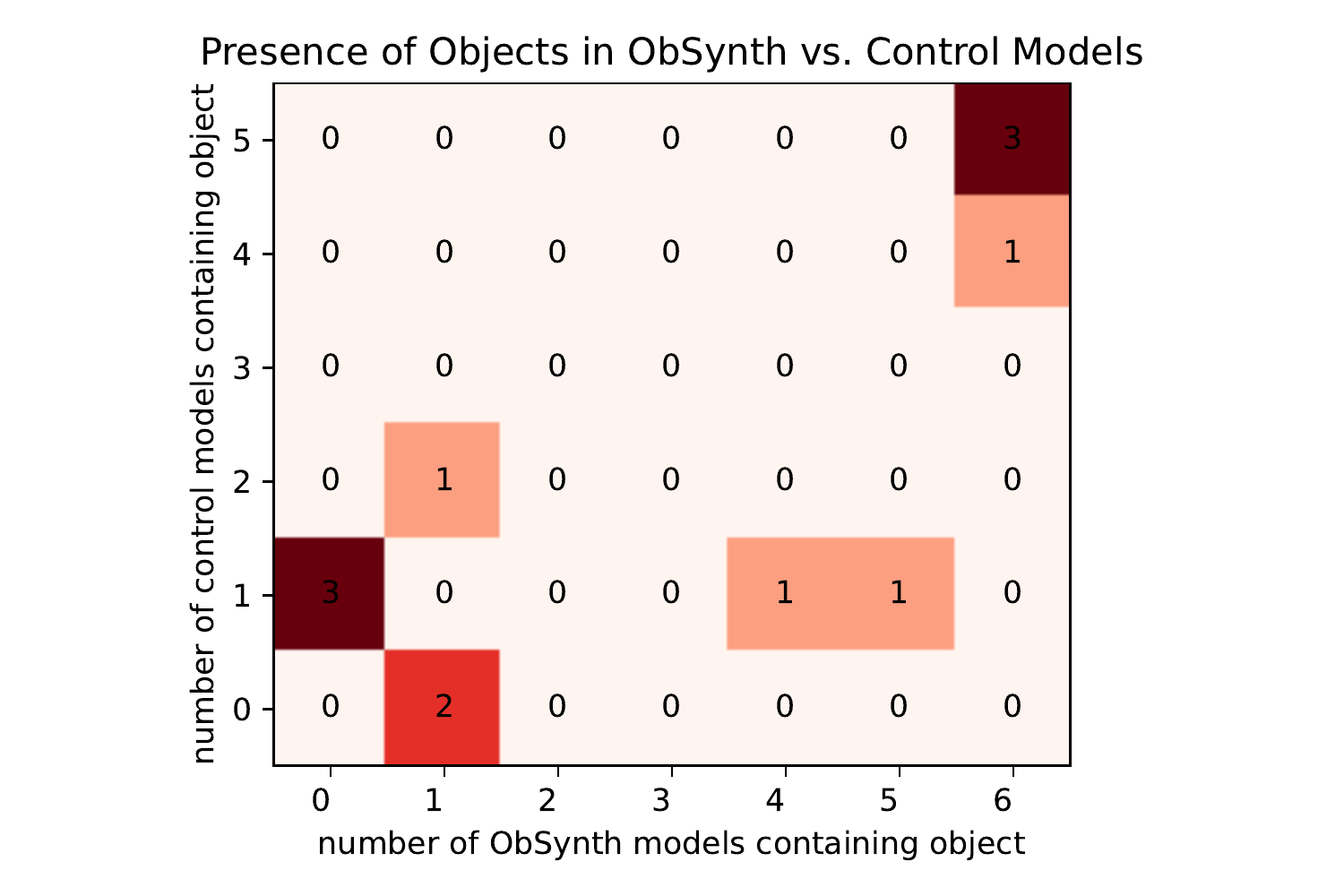}
  \caption{Frequency for objects}
  \label{fig:object-counts}
\end{subfigure}%
\begin{subfigure}{.5\textwidth}
  \centering
  \includegraphics[width=\linewidth]{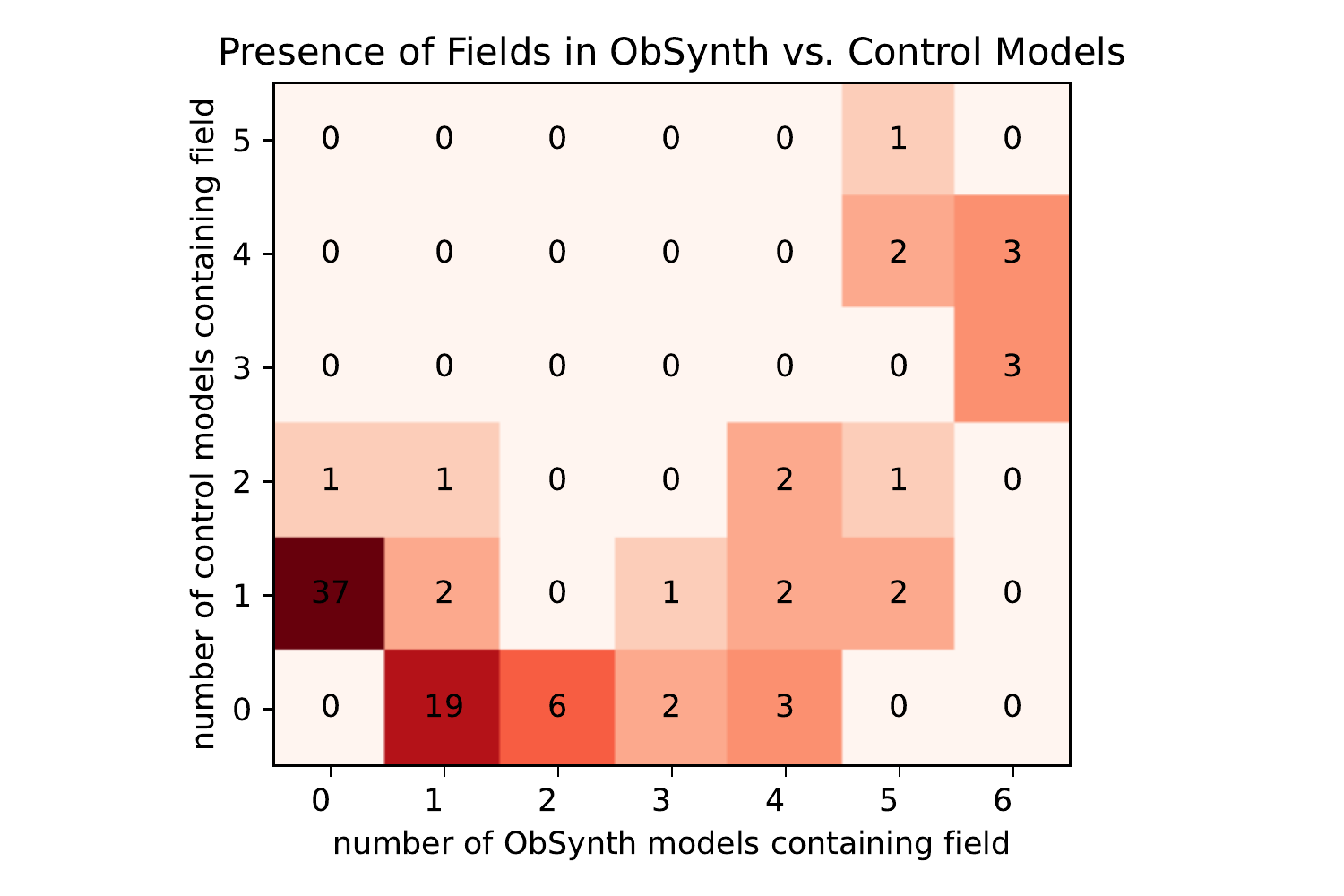}
  \caption{Frequency for fields}
  \label{fig:field-counts}
\end{subfigure}
\caption{This shows how frequently each distinct object (left) and field (right) occurs in both the ObSynth models and the control group models. The lack of objects/fields in the upper left quadrant shows that ObSynth models rarely miss anything that control group models include, while the presence of objects/fields in the lower right quadrant show that ObSynth models consistently include objects/fields that users omitted.}
\label{fig:counts}
\end{figure}

In Fig. \ref{fig:object-counts} we looked at each distinct object name across all object models and we counted the number of times it occurred in models produced by the ObSynth group (x-axis) and the control group (y-axis). We found that the majority of object names were in the top right and bottom left corner, indicating that most objects occurred frequently in both groups or in neither group. The most striking feature of the figure, however is the absence of objects in the upper left quadrant. 
The pattern is even stronger in Fig. \ref{fig:field-counts}, which involves a similar analysis but for field names. Once again, there are no fields in the upper left quadrant, indicating that there are no fields that were common in the control group but rare in the ObSynth group. By contrast, we can see a number of fields in the lower-right quadrant, indicating that many fields that the ObSynth users consistently kept as important were never added by the unaided users. 

To better understand the objects in each of these different quadrants, we looked at the object names themselves (Table \ref{table:obj-counts-qual}). We found that two objects (\textit{waiter} and \textit{table}) were common among ObSynth models but rare in the control group models. Both of these objects made sense in our prompt's context. In addition, the only object common among control group models but rare in ObSynth models was \textit{menu}, but even this object only showed up in 40\% of control group models. This shows that overall, object models produced with the help of ObSynth were more detailed than those produced without.

\begin{table}[!t]
\caption{Qualitative analysis of all distinct objects generated, grouped by how often they occur in the ObSynth models and control group models for a restaurant management application. Note that ObSynth models contain relevant objects (\texttt{waiter}, \texttt{table}) that control group models omit.}
\begin{center}
\begin{tabular}{ |L{0.35\textwidth}|L{0.42\textwidth}| } 
 \hline
 \textbf{Category} & \textbf{Frequencies (\% ObSynth, \% control)} \\ 
 \hline
 high in both & customer, reservation, order (100\%, 100\%); menu item (100\%, 80\%) \\
 \hline
 low in both & chef, restaurant (17\%, 0\%); entree, individual order, drink (0\%, 20\%)\\
 \hline
 high in ObSynth, low in control & waiter (67\%, 20\%); table (83\%, 20\%) \\
 \hline
 low in ObSynth, mid in control & menu (17\%, 40\%) \\
 \hline
\end{tabular}
\end{center}
\label{table:obj-counts-qual}
\end{table}
%
% \begin{table}
% \caption{todo}
% \begin{center}
% \begin{tabular}{ |p{0.35\textwidth}|p{0.42\textwidth}| } 
%  \hline
%  \textbf{Category} & \textbf{Inputs} \\ 
%  \hline
%  only ObSynth & menu item.allergen, table.order, waiter.name \\
%  \hline
%  more in ObSynth & \\
%  \hline
%  similar \\
%  \hline
%  more in control & \\
%  \hline
%  only control & reservation.status \\
%  \hline
% \end{tabular}
% \end{center}
% \label{table:subtasks}
% \end{table}
%
\subsection{(Q2) What is the quality of the objects, fields, and methods initially synthesized by ObSynth before human editing?} 
Next, we wanted to better understand how ObSynth's synthesis capabilities led to the improved completeness of the final models. Were the ObSynth models more detailed because ObSynth suggested fields and objects that did not occur to the unaided users, or was ObSynth simply enabling users to themselves think of additional fields and objects, or in an extreme case, could it be providing suggestions that were so wrong that they forced users to think harder about the problem and generate more detailed object models as a result.

First, looking back to Fig. \ref{fig:object-counts}, we found that the two objects in the bottom-right quadrant were both automatically synthesized by ObSynth in step 2. We saw a similar trend for the fields in the bottom-right quadrant of Fig. \ref{fig:field-counts}: the majority of these were synthesized in step 4. One such example is the \textit{ingredients} field of the \textit{menu item} object, which showed up in 83\% of ObSynth models but only 20\% of control group models. This supports our hypothesis that ObSynth is able to generate fields that users would not have thought of themselves but that they consider useful once they see them. Overall, we found that an average of 92\% of objects, 79\% of fields, and 65\% of methods in the final object models in the ObSynth group were generated in the initial synthesis steps.

We also wanted to understand if objects and fields that ObSynth synthesized remained in the final user-edited model. We first examined the initial object names that ObSynth synthesized in step 2. In all cases, they were \textit{customer, reservation, order, menu item, waiter, table}. These six objects almost always remained in the final ObSynth object models, only \textit{waiter} was removed twice and \textit{table} removed once. Across the six participants in the ObSynth group, we found that an average of 92\% of objects, 73\% of fields and 59\% of methods that were automatically synthesized in step 4 remained in the final object model. This indicates that objects, fields, and methods that ObSynth generated were of decent quality, so the synthesis step aided the user in designing the object model. Qualitatively, participants in this group remarked that the initial objects synthesized in step 2 were ``perfect and complete'' and that synthesis in step 4 ``did most of the work for me'' and ``was accurate most of the time''. Two participants, however, remarked that ``it was sometimes confusing what the generated field names were referring to''. We would like to mitigate this issue in future work by also synthesizing descriptions for each object.

\subsection{(Q3) Do people using ObSynth take less time to design object models?} 
Next, we wanted to understand if ObSynth participants were able to design object models faster than control group participants because of the assisted automation. Measuring the time elapsed from the moment participants clicked \textit{Begin} to \textit{I'm Finished}, we found that the ObSynth participants spent an average of $14.4$ minutes and a standard deviation of $1.6$ minutes, while the control group participants spent an average of $12.0$ minutes with a standard deviation of $3.1$ minutes. Therefore, we saw no evidence that ObSynth helped speed up the task. 

\subsection{(Q4) How does ObSynth shift the workflow of participants building object models?} 
\begin{figure}[]
    \centering
    \includegraphics[width=0.7\textwidth]{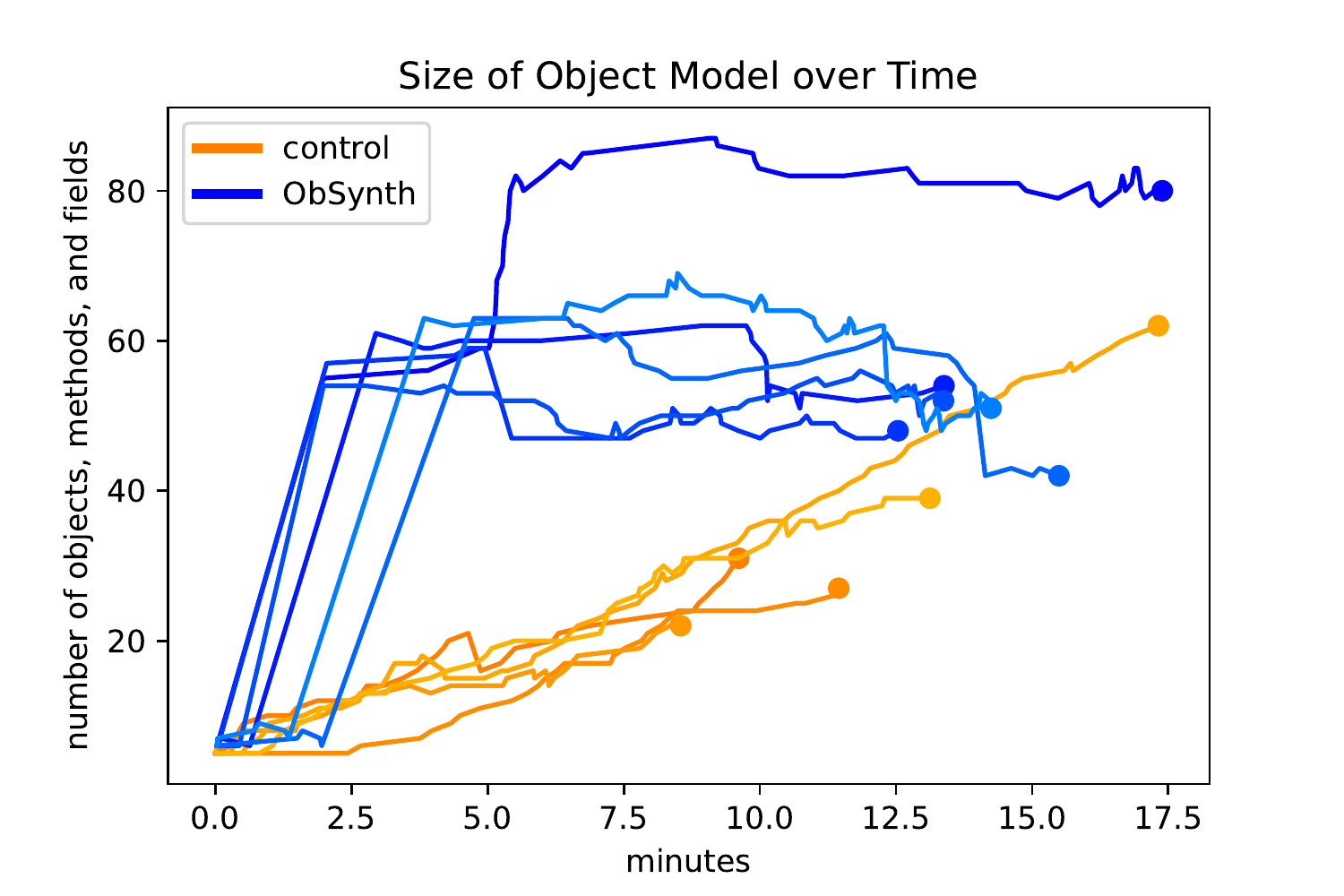}
    \caption{Progress of participants over time for the two groups, measured as the total number of objects, methods, and fields. There is a clear different in the workflow between groups: control group participants incrementally added pieces to their object model, while ObSynth participants took more time examining the synthesized model, as seen by the flat portions.}
    \label{fig:study-progress}
\end{figure}

In order to better understand this phenomenon, we wanted to explore how working with ObSynth shifted the workflow of our participants. Thus, we measured the progress of each participant over time as they designed the object model. As a rough heuristic capturing progress, we plotted the total number of objects, methods, and fields in the object model throughout time in Fig. \ref{fig:study-progress}.
Among the control group, we see a steadily growing trend in progress, representing participants incrementally building up the object model. One participant said, ``I built up the object model step by step, adding things as I thought of them being needed. At the end, I checked to make sure everything was there and consistent''. A different story arises in the ObSynth group: we first see an initial spike in progress, indicating the initial object model generated in step 4. This is followed by a relatively flat period of time where participants are going through the generated model and making corrections. We generally notice a slight decrease in total components from the initial peak, as we designed the synthesis procedure to err on the side of overgeneration. Overall, the total number of components in the final ObSynth models is larger than that for the control group models on average, highlighting again that our tool leads to more detailed models. Participants described their experience working with ObSynth as ``mostly just going through and correcting minor mistakes'' and ``adding a few more methods for convenience and deleting a few things that didn't make sense''.

\section{Conclusion and Future Work}
In this work, we introduced a new class of important program synthesis problems known as \textit{specification reification}, focused on incorporating domain knowledge into traditional program synthesis. We presented one specific instance of this task, object model synthesis, and designed an interactive system, ObSynth, to help users complete this task by automatically synthesizing components of the object model using GPT-3. Through our user study, we showed that ObSynth helps users come up with more detailed object models, generates important pieces of the object model that users often keep, and changes the workflow of users. Importantly, we found the added automation support does not degrade the quality of object models.

We believe that specification reification is an important problem and welcome researchers to introduce other synthesis problems fitting this framework. We identify many attractive directions for future investigation: first, there is significant room for the discovery of better synthesis algorithms for object model generation. Second, we could explore how our system can better understand and incorporate human edits to the object model to synthesize more relevant fields. Third, a useful feature could be to have the model automatically suggest revisions to the object model and catch user mistakes. Finally, it would be interesting to extend the object model to include more features such as method arguments or descriptions of objects, fields, and methods.

\bibliographystyle{ACM-Reference-Format}
\bibliography{sample-base}

\appendix

\section{Prompts for Each Subtask}
In this section, we include the prompts we use to perform each of the eight subtasks in Table \ref{table:subtasks}. We omit (ST1), which is described in the main text.

\textit{(ST2) Synthesize additional object names} This task is used when the user clicks on the \textit{``Auto Add Object"} button in the initial phase. We use the prompt in Fig. \ref{table:initial-tables-more}. This prompt gives the model information on which tables are already present, both for more context and to avoid generating tables that are already present. One note is that we found asking for a specific number of other tables (we used three) helped performance.

\begin{figure}[!ht]
\centering
\begin{tabular}{ |m{0.9\textwidth}| } 
  \hline
  Input: \textit{I want a classroom management app that tracks students, the assignments they've submitted, and the grades they've earned on those assignments. The app has the following tables: student, submission, assignment, teacher.} \\ \\
  \textit{Q: What are three other tables this app might have?} \\
  \textit{A: This app might have the following tables: course, gradebook, attendance.} \\ \\
  \textit{I want a restaurant management app tracking customers, their reservations, their orders, and menu items. The app has the following tables: customer, reservation, order, menu item. } \\ \\
  \textit{Q: What are three other tables this app might have?} \\ \\
  GPT-3 Output: \textit{A: This app might have the following tables: table, dish, price.} \\
  \hline
\end{tabular} 
\caption{GPT-3 prompt to generate additional tables when the user clicks ``Auto Add Object'' in the initial phase}
\label{table:initial-tables-more}
\end{figure}

\textit{(ST3) Synthesizing Field Names:} In this subtask, we take the prompt, the name of an object, and generate a list of fields for that object. The prompt is shown in Fig. \ref{table:prompt-st3}. Of note is the inclusion of two prompts for this task: a classroom management application and a pet store application. We found that with just one prompt, GPT-3 sometimes overfit to that prompt, outputting fields that were classroom application specific despite being fed a completely different prompt.

\begin{figure}[!t]
\centering
\begin{tabular}{ |m{0.9\textwidth}| } 
  \hline
  Input: \textit{I want a classroom management app that tracks students, the assignments they've submitted, and the grades they've earned on those assignments. The app has the following tables: student, submission, assignment, teacher.} \\ \\
  \textit{Q: What does a student have?} \\
  \textit{A: A student has a name, an email address, a list of assignments, a list of submissions, and a teacher.} \\ \\
  \textit{I want a pet store app tracking customers. The pet store app should also keep the store's inventory of pets and allow customers to buy and sell pets.} \\ \\
  \textit{Q: What does a pet have?} \\ \\
  GPT-3 Output: \textit{A: A pet has a name, a species, a breed, a price, a list of customers, and a list of transactions.} \\
  \hline
\end{tabular} 
\caption{GPT-3 prompt to generate candidate from an object name.}
\label{table:prompt-st3}
\end{figure}

\textit{(ST4) Predicting types of fields}: ObSynth uses a very simple prompt engineering approach to predict the type of each field. Our prompt is shown in Fig. \ref{table:type-generation}. Again, we use a primer to demonstrate the type-prediction task. First, we tell GPT-3 the set of potential types, which we extract by including the primitives and the object names in the current table. Then, we give GPT-3 examples of data type prediction questions with correct answers. GPT-3 is able to learn the type prediction task, correctly predicting the data type for the ``price'' attribute in Fig. \ref{table:type-generation}.

\begin{figure}[t]
\centering
\begin{tabular}{ |m{0.9\textwidth}| } 
  \hline
  Input: \textit{I have the following data types: int, float, string, datetime, character, student, teacher, assignment.} \\ \\
  \textit{Q: What is the data type of an attribute named "id"?} \\
  \textit{A: int} \\ \\
  \textit{Q: What is the data type of an attribute named "name"?} \\
  \textit{A: string} \\ \\
  \textit{Q: What is the data type of an attribute named "cost"?} \\
  \textit{A: float} \\ \\
  \textit{Q: What is the data type of an attribute named "time"?} \\
  \textit{A: datetime} \\ \\
  \textit{Q: What is the data type of an attribute named "instructor"?} \\
  \textit{A: teacher} \\ \\
  \textit{I have the following data types: int, float, string, datetime, customer, reservation, order, menu item.} \\ \\
  \textit{Q: What is the data type of an attribute named "price"?} \\
  GPT-3 Output: \textit{A: float} \\ \\
  Input: \textit{Q: What is the data type of an attribute named "items"?} \\
  GPT-3 Output: \textit{A: menu item} \\
  \hline
\end{tabular} 
\caption{GPT-3 prompt to predict data types of fields}
\label{table:type-generation}
\end{figure}

\textit{(ST5) Predicting the multiplicity of a field}: To predict the multiplicity of each field (int vs. List[int]), observe that in Fig. \ref{table:prompt-st3}, we prompted GPT-3 to give us responses of the form \textit{``A student has a name, an email address, a phone number, a list of assignments, and a list of grades.''}. Note that ``a phone number'' is distinguished from ``a list of assignments''. Therefore, if a field is in plural or is preceded by \textit{``list of''}, then ObSynth predicts a list type (shown as ``many'' in the UI). Otherwise, it predicts a base type (shown as ``one''). 

% ST6
% ST6
% ST6
% ST6

\textit{(ST6) Synthesizing method names: } In order to generate methods, ObSynth asks questions like \textit{``Q: What can a pet do?''}, \textit{``Q: What else can a pet do''}, and \textit{``Q: What are the method names for these actions?''}. Then, it asks what the method names for the actions are. The prompt is shown in Fig. \ref{table:prompt-st6}.

\begin{figure}[!ht]
\centering
\begin{tabular}{ |m{0.9\textwidth}| } 
  \hline
Input: \textit{I want a classroom management app that tracks students, the assignments they've submitted, and the grades they've earned on those assignments.} \\
\textit{Q: What can a student do?} \\
\textit{A: A student can view their own submissions, drop a class, or view classroom announcements.} \\ 
\textit{Q: What else can a student do?} \\ 
\textit{A: A student can submit an assignment or view their grades.} \\
\textit{Q: What are the method names for these actions?} \\
\textit{A: The method names are: viewSubmissions, dropClass, viewAnnouncements, submitAssignment, viewGrades} \\ \\
\textit{I want a pet store app tracking customers. The pet store app should also keep the store's inventory of pets and allow customers to buy and sell pets.} \\
\textit{Q: What can a pet do?} \\
GPT-3 Output: \textit{A: A pet can be bought or returned.} \\ \\
Input: \textit{Q: What else can a pet do?} \\
GPT-3 Output: \textit{A: A pet can have its information updated.} \\ \\
Input: \textit{Q: What are the method names for these actions?} \\
GPT-3 Output: \textit{A: The method names are: buyPet, sellPet, updateInformation} \\
\hline
\end{tabular} 
\caption{GPT-3 prompt for generating method names}
\label{table:prompt-st6}
\end{figure}

\textit{(ST7) Synthesize additional field names}: This subtask is used when the user clicks \textit{Auto Add Field}. The prompt is shown in Fig. \ref{table:prompt-st7}.
\begin{figure}[!ht]
\centering
\begin{tabular}{ |m{0.9\textwidth}| } 
  \hline
Input: \textit{I want a classroom management app that tracks students, the assignments they've submitted, and the grades they've earned on those assignments.} \\
\textit{Q: What are 3 other things a student might have?} \\
\textit{A: A student might have a list of submissions, an address, and a list of assignments.} \\ \\
\textit{I want a task management application where users can create and manage tasks, set deadlines, and receive notifications.A user has and a list of notifications. A task has a name, a description, a deadline, a list of subtasks, and a list of users. A deadline has a name, a description, a date, and a list of tasks.} \\
\textit{Q: What are 3 other things a user might have?} \\
GPT-3 Output: \textit{A: A user might have a profile, a list of tasks, and a list of notifications.
} \\
\hline
\end{tabular} 
\caption{GPT-3 prompt for generating additional field names}
\label{table:prompt-st7}
\end{figure}

\textit{(ST8) Synthesize additional method names}: This subtask is used when the user clicks \textit{Auto Add Method}. The prompt is shown in Fig. \ref{table:prompt-st8}.
\begin{figure}[!ht]
\centering
\begin{tabular}{ |m{0.9\textwidth}| } 
  \hline
Input: \textit{I want a classroom management app that tracks students, the assignments they've submitted, and the grades they've earned on those assignments.} \\
\textit{The student object has the following methods: submitAssignment(), seeGrades(), addCourse().} \\
\textit{Q: What other methods does the student object have?} \\
\textit{A: getAssignments(), editSubmission()} \\ \\
\textit{I want a library app to track books. The customers can borrow books and return books from the library.} \\
\textit{The customer object has the following methods: borrowBook(), searchBook(), returnBook().} \\
\textit{Q: What other methods does the customer object have?} \\
GPT-3 Output: \textit{A: checkoutBook(), reserveBook(), renewBook()} \\
\hline
\end{tabular} 
\caption{GPT-3 prompt for generating additional method names}
\label{table:prompt-st8}
\end{figure}

\section{Qualitative Results for User Study}
\subsection{Generated Object Models}
\begin{figure}[!ht]
    \includegraphics[width=0.9\textwidth]{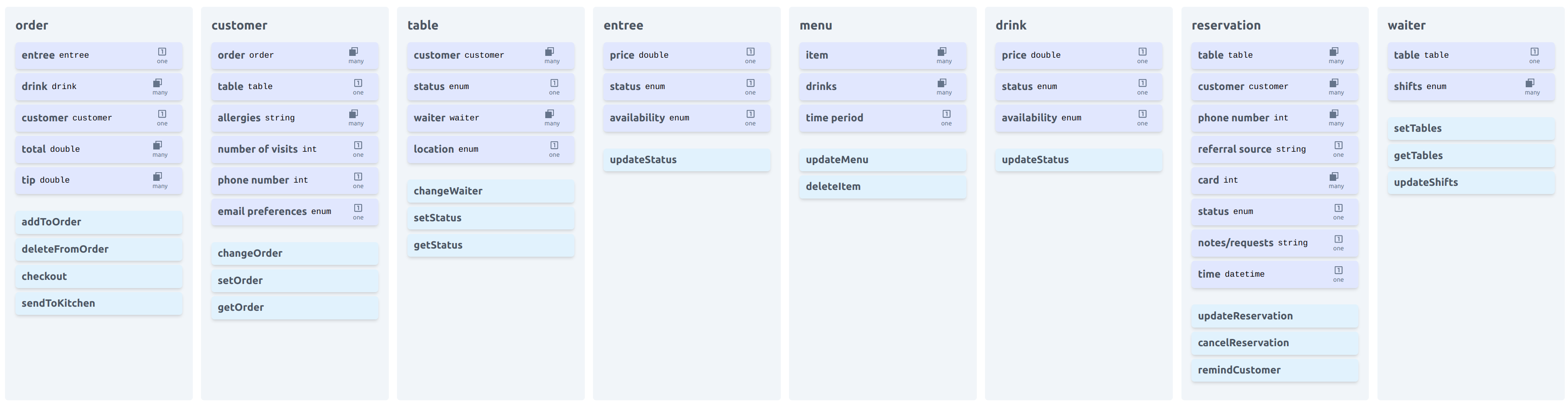}\hfill
    \includegraphics[width=0.9\textwidth]{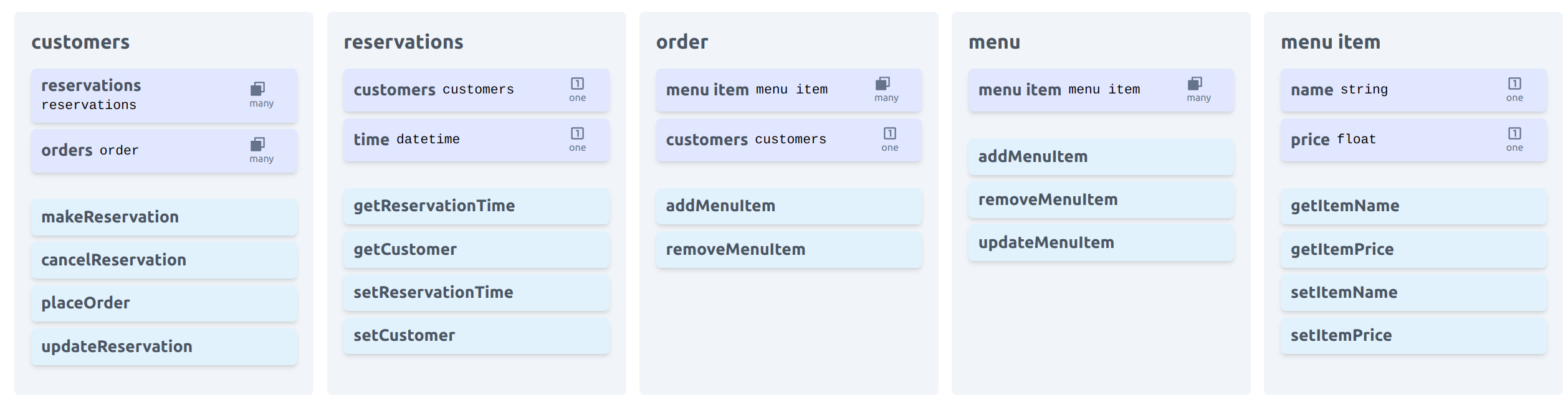}\hfill
    \includegraphics[width=0.9\textwidth]{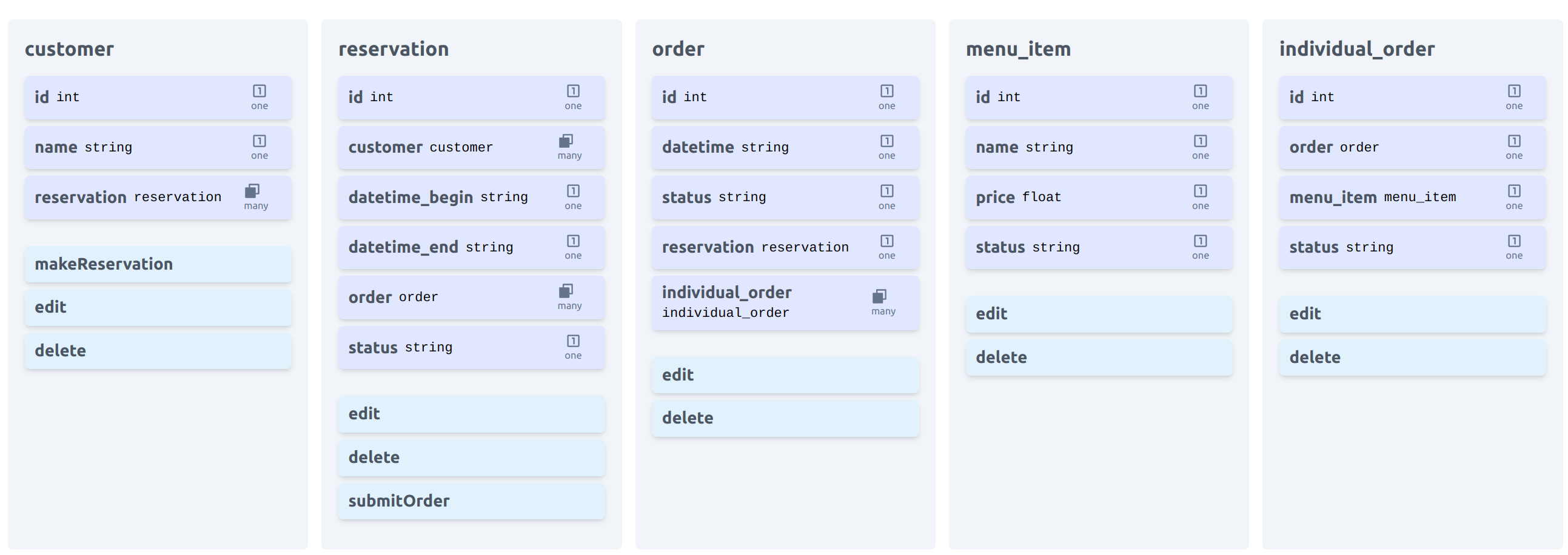}\hfill
    \includegraphics[width=0.9\textwidth]{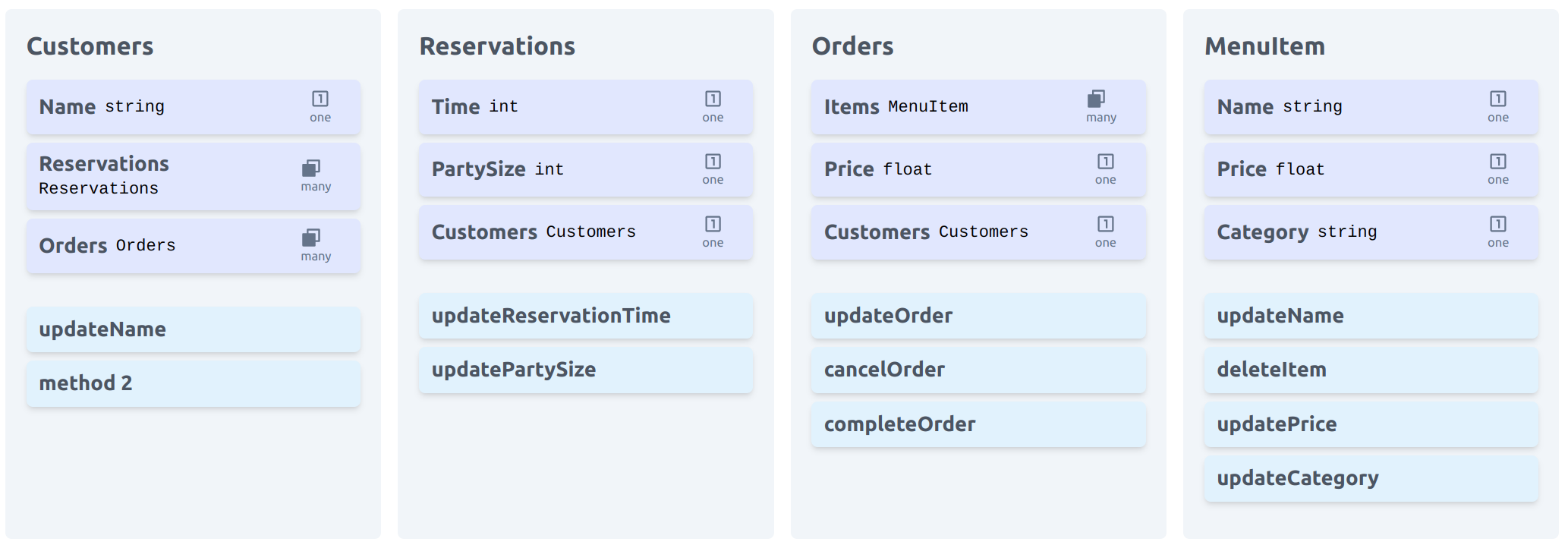}\hfill
    \includegraphics[width=0.9\textwidth]{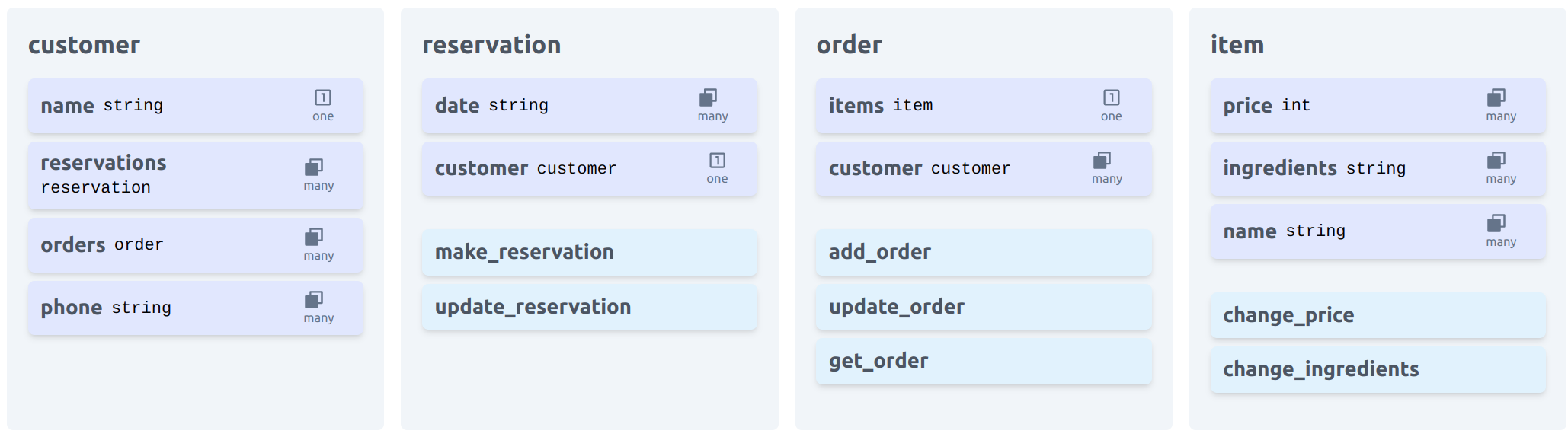}\hfill
    \caption{Object models designed by users in the control group for a restaurant management application}\label{fig:schemas-nogpt}
\end{figure}

We include all the object models generated by the 11 participants in our study. The object models from the six users using ObSynth are shown in Fig. \ref{fig:schemas-gpt}, while those from the five models without ObSynth are shown in Fig. \ref{fig:schemas-nogpt}.

\begin{figure}[!ht]
    \includegraphics[width=\textwidth]{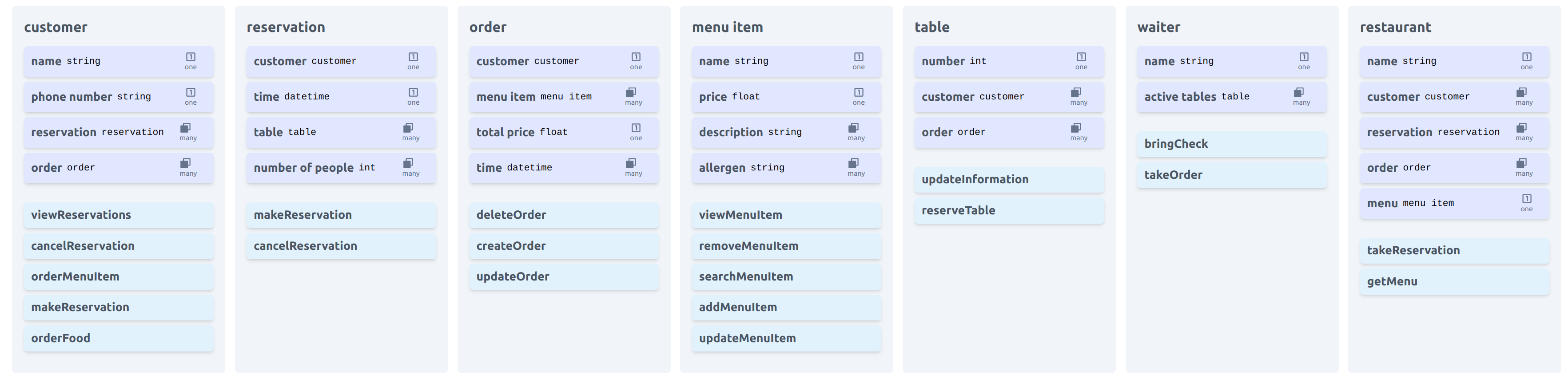}\hfill
    \includegraphics[width=\textwidth]{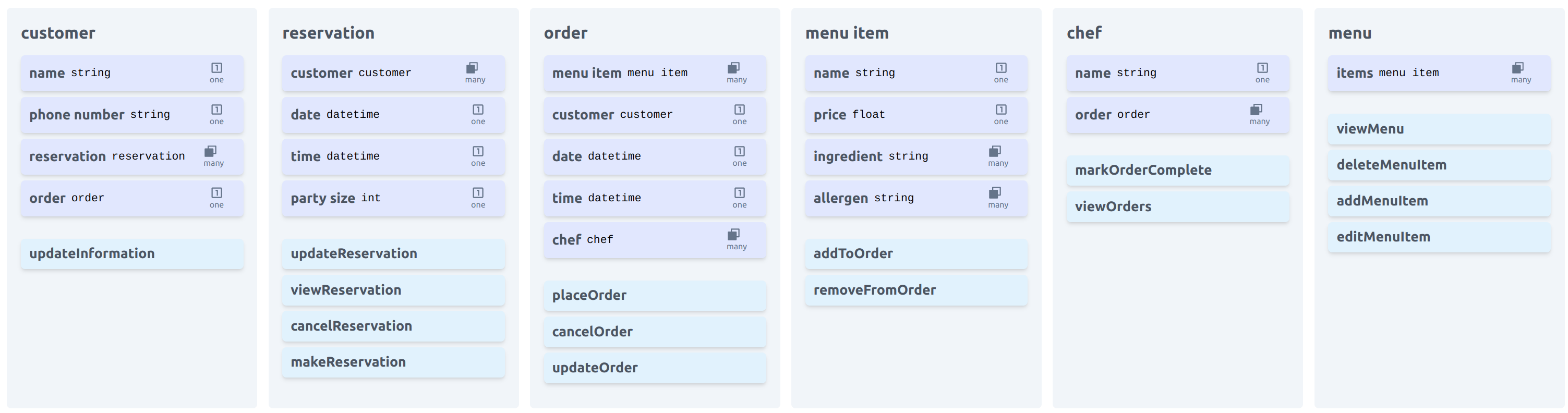}\hfill
    \includegraphics[width=\textwidth]{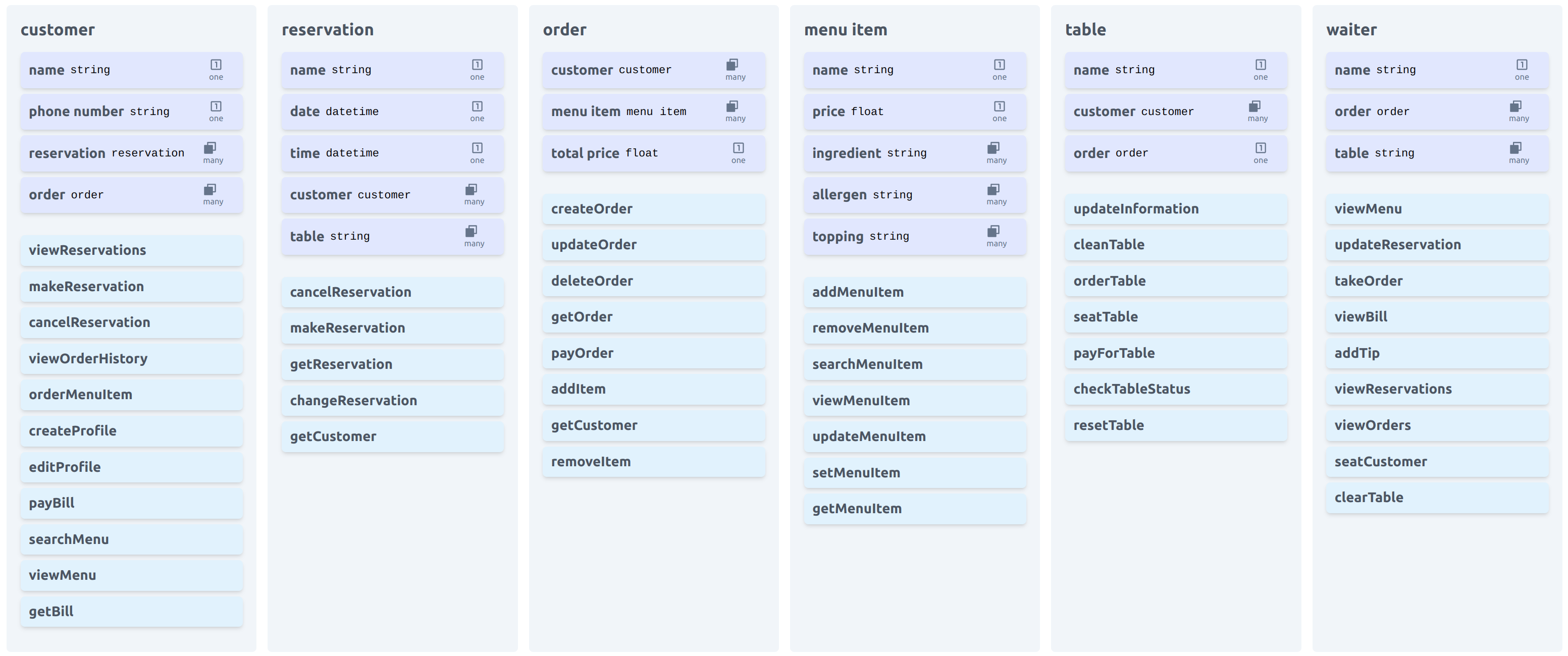}\hfill
    \caption{Object models designed by users with the help of ObSynth for a restaurant management application}
\end{figure}
\begin{figure}[!ht]
    \ContinuedFloat
    \includegraphics[width=\textwidth]{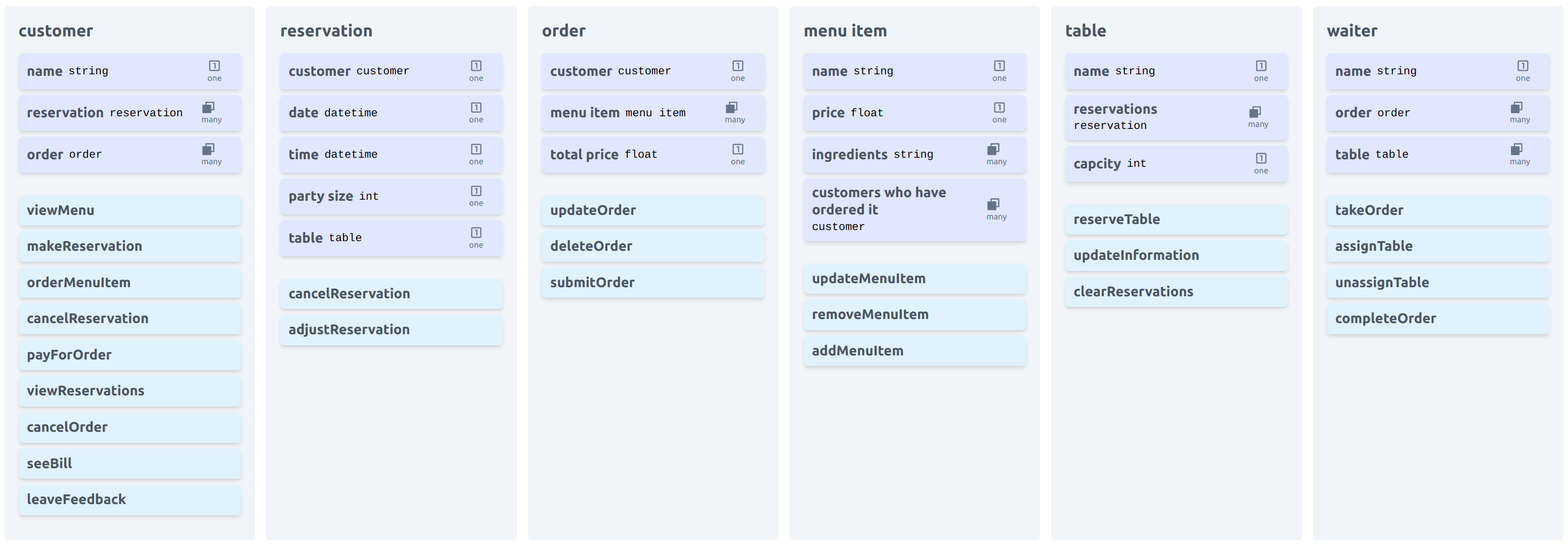}\hfill
    \includegraphics[width=\textwidth]{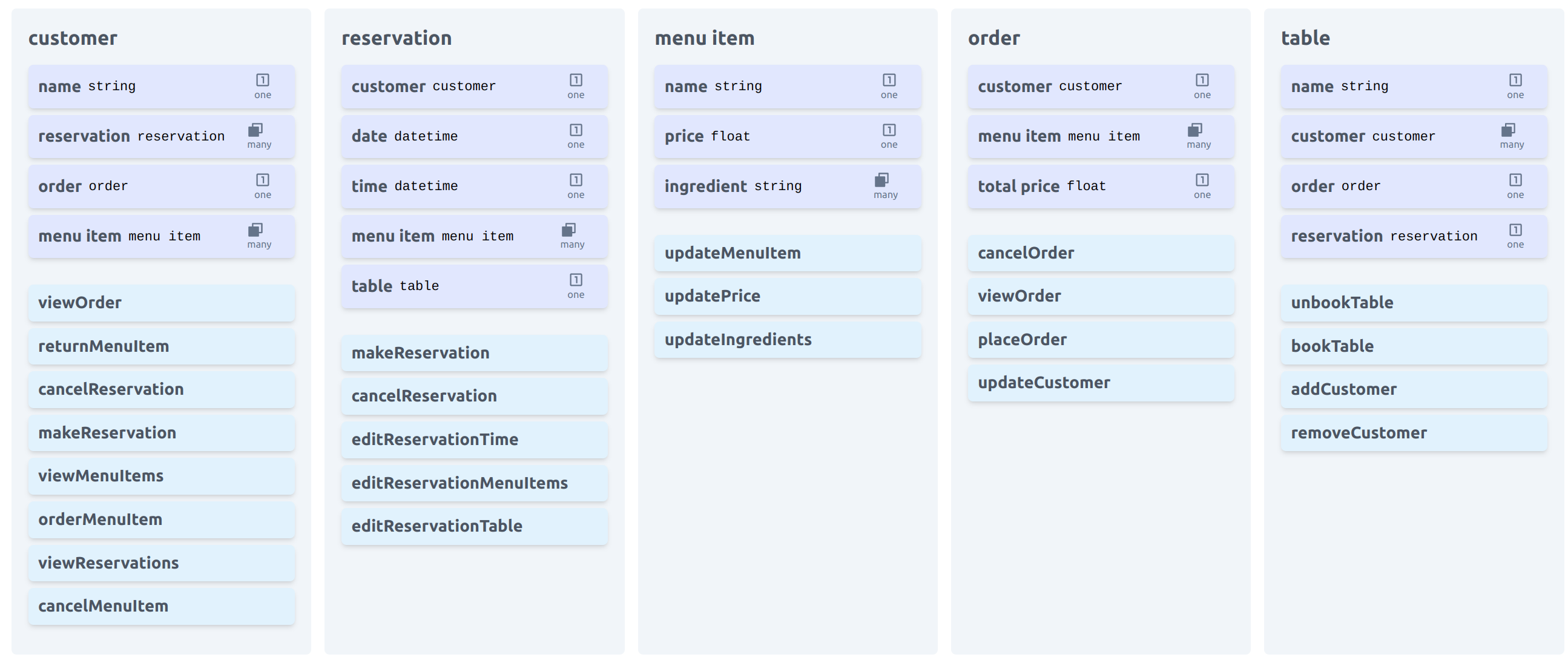}\hfill
    \includegraphics[width=\textwidth]{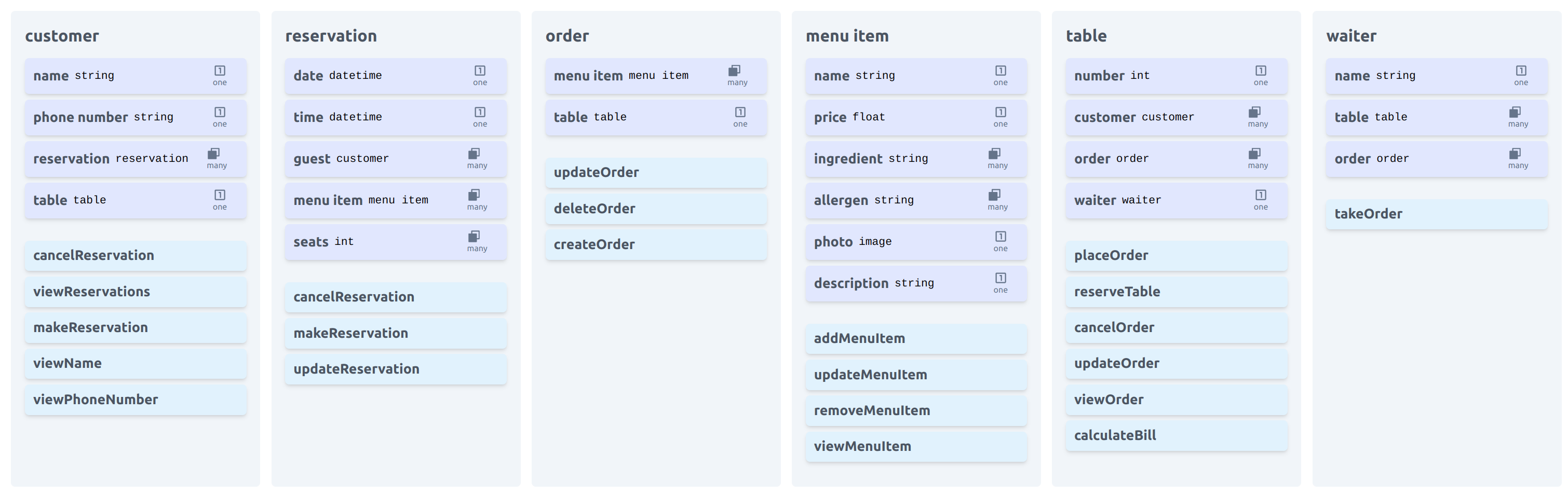}\hfill
    \caption{Object models designed by users with the help of ObSynth for a restaurant management application, cont. }
    \label{fig:schemas-gpt}
\end{figure}

\clearpage
\newpage
\subsection{Feedback}
\textit{ObSynth group feedback}: Participants in the ObSynth group found that the suggestions were of decent quality, stating:
\begin{itemize}
\item ``The initial suggestions for the objects was very spot on''
\item ``The fields themselves were pretty intuitive, I didn't have to make many edits''
\item ``It was able to understand some of the concepts around the quantity of things and the types.''
\item ``A lot of the adjustments I had to make were from my initial design decision of not taking orders from the kitchen''
\item ``I think definitely the suggestions were helpful, it was a matter of deciding what I wanted to keep in my particular case''
\end{itemize}

In addition, one participant commented on the difference in workflow: ``It's always easier to react to something than create something from scratch. You can basically build CRUD apps very quickly using something like this.'' On the other hand, two participants remarked that ``The methods were a bit worse, it had a better time predicting fields that were useful'' and that ``it was sometimes confusing what the generated field names were referring to,'' indicating that ObSynth still occasionally made mistakes while generating and that human feedback was necessary to correct the mistakes. 

Participants in the ObSynth group also provided feedback on the task setup and UI, saying:
\begin{itemize}
\item ``It would have been helpful to have a diagram which shows the relationship between different objects''
\item ``I found it confusing that some of the [methods] were multi-table operations, but just under one table''
\item ``One-way/two-way was confusing, understanding it as a many-many relationship would have been helpful''
\end{itemize}

Finally, one participant commented that ``I would be curious to see how it did in a slightly more complicated example, since this one is probably very common,'' which is a direction of future work. \newline

\noindent\textit{Control group feedback}: Users in the control group commented mostly on the UI and general experience, as they did not have access to ObSynth's automation features. There were many positive comments, indicating that even just having a UI is helpful for this task:
\begin{itemize}
\item ``It was positive, overall a good experience''
\item ``It was really nice, I liked the general aesthetic''
\item ``I think the tool is pretty easy to use, I liked being able to add, edit, and delete with icons''
\item ``I like that it autopopulated when you created a connection, one way or many''
\end{itemize}

We also received helpful constructive feedback on our task setup:
\begin{itemize}
\item ``There were a few UI things, sometimes the fields overlapped and ran into each other''
\item ``I noticed myself clicking on the fields instead of the green checkmark''
\item ``Maybe offer more relational tools, show how the different objects related to each other graphically''
\end{itemize}

Finally, a few participants asked for more automation:
\begin{itemize}
\item ``Maybe all the CRUD operations could have been added by default''
\item ``I wish there were a dropdown menu [or autocompletion] for types''
\item ``I don't know if this is possible, but I wish there were a way to pre-populate fields, because it's very easy to mess up typing''
\end{itemize}

\end{document}